\documentclass[useAMS,usenatbib]{mn2e}

\def\figs{ANGMOM_FIGS}
\usepackage{graphicx}
\usepackage{color}
\usepackage{rotating}
\usepackage{pdflscape}
\usepackage{float}

\title[Accretion-regulated angular momentum evolution]
  {Accretion discs as regulators of stellar angular momentum evolution in the ONC and Taurus-Auriga}
\author[C. L. Davies et al.]
  {Claire L. Davies$^1$\thanks{E-mail: cd54@st-andrews.ac.uk}, Scott G. Gregory$^1$ and Jane S. Greaves$^1$\\
  $^1$SUPA School of Physics and Astronomy, University of St Andrews, North Haugh, St Andrews, Fife KY16 9SS\\}

\date{2014 July 23}

\pagerange{\pageref{firstpage}--\pageref{lastpage}}\pubyear{2014}

\begin{document}

\label{firstpage}

\maketitle

\begin{abstract}
In light of recent substantial updates to spectral type estimations and newly established intrinsic colours, effective temperatures, and bolometric corrections for pre-main sequence (PMS) stars, we re-address the theory of accretion-disc regulated stellar angular momentum (AM) evolution. We report on the compilation of a consistent sample of fully convective stars within two of the most well-studied and youngest, nearby regions of star formation: the Orion Nebula Cluster (ONC) and Taurus-Auriga. We calculate the average specific stellar AM ($j_{\star}$) assuming solid body rotation, using surface rotation periods gathered from the literature and new estimates of stellar radii and ages. We use published $\itl{Spitzer}$ IRAC fluxes to classify our stars as Class II or Class III and compare their $j_{\star}$ evolution. Our results suggest that disc dispersal is a rapid process that occurs at a variety of ages. We find a consistent $j_{\star}$ reduction rate between the Class II and Class III PMS stars which we interpret as indicating a period of accretion disc-regulated AM evolution followed by near-constant AM evolution once the disc has dissipated. Furthermore, assuming our observed spread in stellar ages is real, we find the removal rate of $j_{\star}$ during the Class II phase is more rapid than expected by contraction at constant stellar rotation rate. A much more efficient process of AM removal must exist, most likely in the form of an accretion-driven stellar wind or other outflow from the star-disc interaction region or extended disc surface.  
\end{abstract}


\section{Introduction}
If all angular momentum (AM) was conserved during contraction from a natal molecular cloud to the zero-age main sequence (ZAMS), stellar rotational velocities would far exceed those required to break a star apart. A solar mass star, accreting at a typical rate of $10^{-7}\,\rmn{M_{\odot}yr^{-1}}$ would reach its break-up velocity after just $1\,$Myr \citep{Hartmann1989}. However, stars with accretion discs are found to be rotating at much slower rates, suggesting that significant AM removal mechanisms must operate during the first few Myr of formation \citep{Bouvier1986, Hartmann1986}. 

Accretion disc-regulated AM removal was initially attributed to a magnetic torque produced by the differential rotation between a star and its Keplerian disc \citep{Ghosh1979, Camenzind1990, Konigl1991, Collier1993}. For this torque to sufficiently brake the star, the stellar magnetic field would need to interact with a region in the disc beyond the corotation radius and be stable over multiple rotations. However, differential twisting of the magnetic field lines, together with the competing processes of accretion and diffusion, limit the size of the connected region in the disc and reduce the extent of the field beyond corotation \citep{Shu1994, Bardou1996, Agapitou2000, Matt2005, Zanni2009}. Thus, such a mechanism would be insufficient to spin down the star. These findings have prompted more recent theoretical studies to favour star-disc interaction related magnetised winds and outflows as possible AM removal mechanisms in actively accreting pre-main sequence (PMS) stars \citep{Shu1994, Lovelace1995, Matt2005, Zanni2013}. 

Observational studies of AM evolution in PMS stars primarily focussed on the distribution of stellar surface rotation rates. Until the formation of a radiative core, stellar rotation can be approximated to that of a solid body. Therefore, while the PMS star is fully convective, the surface rotation period can be used to study the AM of the entire star. In young star forming regions such as the Orion Nebula Cluster (ONC), NGC~2264, IC~348, and Taurus-Auriga, distributions of PMS surface rotation periods were observed to be bimodal (e.g. \citealt{Attridge1992, Edwards1993, Choi1996, Herbst2000}; \citealt*{Cohen2004}; \citealt{Herbst2005, Lamm2005, CiezaBaliber2007}) with the peak of slower rotators interpreted as indicating disc-regulated AM removal. Once the disc dissipates, the star conserves AM, spinning up as it contracts, and is observed in the peak of more rapid rotators. 

Accretion disc-regulated PMS AM evolution has not found unanimous support. Certain studies have not observed a relationship between stellar rotation and accretion disc indicators. However, these contrasting findings can be explained in terms of a variety of biases, masking the underlying relationship between accretion and rotation. For instance, early studies of PMS rotation rates were affected by aliasing and beat phenomena (e.g. \citealt{Stassun1999}), inclusion of non-members (e.g. \citealt{Rebull2001}), and unreliable indicators of accretion discs (e.g. \citealt{Makidon2004}). Furthermore, an underlying relationship between rotation rate and stellar mass has been uncovered \citep{Herbst2002, CiezaBaliber2007}, used to explain the more unimodal rotation period distributions seen in some studies (e.g. \citealt{Stassun1999}). This mass effect is partially attributable to the comparative sizes of ``high'' and ``low'' mass stars. For a sample of stars of a given age and specific stellar AM, $j_{\star}$, those with lower masses will have smaller radii, $R_{\star}$. Since $j_{\star}\propto R_{\star}^{2}/P$, the rotation periods, $P$, of the lower mass sample will be shorter than the higher mass sample \citep*{Herbst2001}. Thus, the lower mass slow rotators are shifted towards the peak of rapid rotators, blurring the bimodality found for the higher mass sample. 

\citet{CiezaBaliber2007} found the bimodality of the rotation period distributions to be severely affected by even a small contamination of stars with spectral types later than M2. The difference in size between the higher and lower mass stars cannot explain this alone and the location of this boundary remains poorly understood. The most promising underlying physical explanation relates to changes in the strength and geometry of the large-scale stellar magnetic field around this spectral type \citep{Lamm2005}. 

The efficiency of AM removal via magnetised winds or outflows is related to the relative strength of the dipole component of the magnetic field as this governs the position of the disc truncation radius and the level of flux from open magnetic fields \citep{Gregory2008, Adams2012, Johnstone2014}. \citet{Donati2011a} found this mechanism to be most efficient in PMS stars of $\sim 0.5-1.3\,\rmn{M_{\odot}}$. The growth of a radiative core in higher mass stars inhibits the build up of a strong dipole field \citep{Donati2011a} and lower mass stars, although still fully convective, appear to have weaker large-scale magnetic fields \citep{Donati2010a, Gregory2012, Donati2013}. Thus, the magnetic fields of stars later than M2 truncate their discs closer to the star, meaning they rotate more rapidly than their higher mass, fully convective counterparts. Although less efficient for lower mass stars, accretion disc regulation can still explain their AM evolution during the first few Myrs (\citealt*{Rodriguez2010}; \citealt{Irwin2011}).

In this paper we focus on the evolution of specific stellar AM ($j_{\star}$) in two of the youngest, nearby regions of star formation, namely the ONC ($\sim1\,$Myrs; \citealt{Hillenbrand1997}) and Taurus-Auriga ($\sim2.8\,$Myrs; \citealt{White2001}). The well-studied nature and youthful ages of these two regions allows us to split the sample according to their position in the Hertzsprung-Russel (HR) diagram into fully convective and partially convective samples. In Section \ref{AMtheory}, we summarise the model used to calculate $j_{\star}$ and how we determined which stars are fully convective. Section \ref{AMdata} details how the data required to calculate $j_{\star}$ was obtained and how we split our data into ``high mass'' and ``low mass'' samples. Our results are detailed in Section \ref{results} and summarised in Section \ref{summary}. 

\section{Angular momentum model}\label{AMtheory}
The AM of a rotating object is a product of its moment of inertia and angular velocity. Thus, in order to calculate the stellar AM, we need to be able to model the distribution and rotation of stellar material. This calculation is greatly simplified for low mass PMS stars as they are fully convective during at least the first few Myrs of contraction \citep{Limber1958, Chabrier1997, Gregory2012}. Thus, they lack the layer of high rotational velocity sheer that exists at the boundary between the radiative core and convective envelope in partially convective stars like the Sun, and generates surface differential rotation.

Studying the levels of differential rotation present on stellar surfaces is possible with tomographic Doppler imaging techniques. This requires spectroscopic monitoring of stars over a few rotations and with sufficient phase coverage. As this is telescope time intensive, to date the surface differential rotation rate, $\mathrm{d}\Omega$, has only been measured for a handful of fully convective PMS stars. The fully convective, non-accreting PMS stars, TWA~6, LkCa~4, and V410~Tau each have differential rotation rates consistent with solid body rotation ($\mathrm{d}\Omega=0$) to within $1.7\,\sigma$ \citep[submitted]{Skelly2008, Skelly2010, Carroll2012, Donati2014}. However, \citet{Donati2010a} found that the fully convective accreting PMS star V2247~Oph exhibited substantial differential rotation with $\mathrm{d}\Omega=0.32\pm0.05\,\mathrm{rad}\,\mathrm{day^{-1}}$. To date, this is the only fully convective PMS star with measured surface differential rotation. It is also the lowest mass star of the sample and has a large-scale magnetic field that is more complex than that of higher mass fully convective PMS stars. It may exist in a regime of dynamo bistability, whereby stars with otherwise similar parameters have drastically different large-scale magnetic topologies and surface differential rates, as has been found for the lowest mass main-sequence M-dwarfs (c.f. \citealt{Gregory2012}). Further observations are required to determine how common differential rotation like that observed in V2247~Oph is.

For now, we assume that the surfaces of fully convective PMS stars rotate as solid bodies (the usual assumption of stellar evolution models e.g. \citealt{Eggenberger2012}). This is consistent with the observations of TWA~6, LkCa~4 and V410~Tau, as well as the observational study of \citet{Barnes2005} who found a decrease in surface differential rotation with increasing convective zone depth. Solid body rotation is also typically found in numerical models (e.g. \citealt{Kuker1997}) and magneto-hydrodynamic simulations (e.g. \citealt{Browning2008}) of fully convective stars. 

For a fully convective star of mass, $M_{\star}$, and radius, $R_{\star}$, rotating with angular velocity, $\Omega=2\pi/P$, the specific stellar AM is then given by
\begin{equation}\label{Jstar}
	j_{\star}=\frac{J_{\star}}{M_{\star}}=\frac{2\pi k^{2} R_{\star}^{2}}{P},
\end{equation}
 where $J_{\star}$ is the stellar AM, $P$ is the rotation period, and $k$ is the radius of gyration \citep{Chandrasekhar1950, Krishnamurthi1997, Herbst2005}. For a perfect sphere, $k^{2}=(2/3)$. However, the most rapidly rotating stars in our sample will be distorted from a spherical shape. To account for this, we explicitly calculate the radius of gyration for each individual star. Following \citet{Herbst2005},
\begin{eqnarray}
	k^{2} & = & \left( \frac{4}{3}a^4 + \frac{16}{15}a^3 b + \frac{8}{7} a^2 b^2 + \frac{16}{105} a b^3 + \frac{52}{1155}b^4 \right) \\
	\nonumber & & \times \left( 2a^2 + \frac{2}{5}b^2 \right)^{-1} 
\end{eqnarray}
 where, if we model a PMS star as a polytrope of index, $n=3/2$,
\begin{equation}\label{a}
	a=1.74225\nu + 1
\end{equation}
 and
\begin{equation}\label{b}
	b=3.86184\nu.
\end{equation}
 Here,
\begin{equation}
	\nu = \frac{2\pi}{G P^2 \rho_{\rmn{c}}}
\end{equation}
 where $G$ is the gravitational constant and $\rho_{\rmn{c}}$ is the central density of the star \citep{Chandrasekhar1935}.

Equation (\ref{Jstar}) is valid for each star until it forms a radiative core. The age at which this happens is dependent on the stellar mass. Stars below $\sim0.35\,\rmn{M_\odot}$ remain fully convective throughout their formation and during their main sequence (MS) lifetimes \citep{Limber1958, Chabrier1997}. More massive stars form radiative cores during their contraction. \citet{Gregory2012} derived the mass-dependent age at which a star of mass greater than $0.35\,M_{\odot}$ would develop a radiative core where
\begin{equation}\label{tcore}
	t_{\rmn{core}}\approx\left(1.494\frac{M_{\odot}}{M_{\star}}\right)^{2.364}\,\rmn{Myrs},
\end{equation}
 based on \citet*{Siess2000} PMS evolutionary models. Accordingly, we limit our analysis to stars with isochronal ages (Section \ref{massage}) below their individual $t_{\rmn{core}}$ limit. 

\section{Stellar data}\label{AMdata}
In order to calculate $j_{\star}$ using equation (\ref{Jstar}) and study its evolution for fully convective stars in the ONC and Taurus-Auriga, we required estimates of stellar masses, radii, central densities, ages, and rotation rates as well as reliable indicators of accretion disc presence. Both the ONC and Taurus-Auriga have well studied stellar populations (e.g. \citealt{Hillenbrand1997, Luhman2010}) but a range of different methods have been adopted to calculate their properties. In the majority of cases, estimates of stellar masses and ages have relied on the comparison of observationally derived effective temperatures, $T_{\rmn{eff}}$, and bolometric luminosities, $L_{\star}$, or colour-magnitude diagrams, to theoretical PMS evolutionary models. However, the choice of PMS evolutionary model differs between studies and multiple methods have been employed to translate spectral types and optical magnitudes into $T_{\rmn{eff}}$ and $L_{\star}$. 

To compare our findings for the ONC with those of Taurus-Auriga, it was necessary to assign $T_{\rmn{eff}}$ and calculate $L_{\star}$ in a fully consistent manner. We gathered spectral types and optical magnitudes from the literature, as detailed below. We adopt the recently derived scales of \citet{Pecaut2013} which account for the bluer colours of PMS stars by accounting for the combined effects of their lower surface gravities \citep{Luhman1999, Dario2010, Pecaut2013, Herczeg2014} and spotted surfaces \citep{Gullbring1998, Stauffer2003} and are thus more applicable here than typically used MS dwarf scales (e.g. \citealt{Bessell1988, Bessell1995, Kenyon1995, Luhman1999}). Details of this process are presented in Sections \ref{SpT_Teff_section} and \ref{Lbol_section}. We calculate stellar radii using our values of $T_{\rmn{eff}}$ and $L_{\star}$ and adopt the \citet{Siess2000} PMS evolutionary models to translate $T_{\rmn{eff}}$ and $L_{\star}$ into stellar masses and ages. Details of these processes are outlined in Sections \ref{massage} and \ref{radiispread}.

For the stellar rotation rates, we gathered previously determined rotation periods from the literature. By using rotation periods rather than projected rotational velocities, $v\sin i$, we removed the dependence on unknown stellar inclinations. The sources of rotational period data used, as well as the checks we employed to ensure we avoided previously reported sources of bias, are presented in Section \ref{Prot_section}. 

To study the dependence of AM evolution on the presence of an accretion disc, we identified all Class II and Class III PMS stars in our ONC and Taurus-Auriga samples. The details of this process are outlined in Section \ref{disc_diagnosis}. 

The compiled datasets for the ONC and Taurus-Auriga are presented in Tables \ref{ONCtable} and \ref{TAUtable}, respectively. These tables include all members (Section \ref{membership}) for which a spectral type was available that had not previously been identified as a binary or multiple system (Section \ref{multiplicity}). Stars found not to be fully convective (Section \ref{AMtheory}) were removed from the analysis but are included in Tables \ref{ONCtable} and \ref{TAUtable} for completeness. 

\subsection{Effective temperatures}\label{SpT_Teff_section}
Spectroscopically determined spectral types were gathered from the literature. Tables \ref{ONCtable} and \ref{TAUtable} list the individual reference for each star in our ONC and Taurus-Auriga samples, respectively. For the bulk of ONC stars, spectral types were retrieved from the newly updated cluster census of \citet*{Hillenbrand2013}. Where multiple spectral types were retrieved for the same star, good agreement was found in general but, in the instances where studies had determined different spectral types, preference was given to the most recent studies. 

A number of very low-mass stars in the ONC did not have spectroscopically-determined spectral types available in the literature. However, some of these did have spectral types calculated using the $7770\,${\AA\/} narrow-band filter in \citet{Dario2010}. In a recent study, \citet{Hillenbrand2013} found a seemingly large scatter between these photometrically-determined spectral types and those determined spectroscopically. However, they also noted that their newly determined spectral types for the lowest mass stars also displayed a similar level of scatter compared to previous spectral types. With this in mind, we adopt these photometrically-determined spectral types for the very low-mass stars with no spectroscopically-determined spectral types. 

We made use of newly derived spectral type-to-$T_{\rmn{eff}}$ conversions for $5$--$30\,$Myr old PMS stars detailed in table 6 of \citet{Pecaut2013}. Although both the ONC and Taurus-Auriga are younger than $5\,$Myrs, these effective temperatures are more applicable than the typically used MS dwarf scales (e.g. \citealt{Bessell1988, Bessell1995, Kenyon1995, Luhman1999}) as they take into account the lower surface gravities and the presence of cool starspots on the surfaces of PMS stars \citep{Gullbring1998, Luhman1999, Stauffer2003, Dario2010, Pecaut2013, Herczeg2014}. However, they are only available for stars of spectral type F0 to M5. This has little effect on our results as most stars later than M5 have masses below $0.1\,\rmn{M_\odot}$ and therefore fall below the lowest mass track in the \citet{Siess2000} models (which we adopt to estimate stellar masses and ages, see Section \ref{massage}) and stars of spectral types earlier than F0 are too massive to be T Tauri stars. For the seven stars in our sample with spectral types later than M5, the spectral type-to-$T_{\rmn{eff}}$ conversions for MS dwarfs detailed in table 5 of \citet{Pecaut2013} were adopted for continuity. 

Where individual errors on spectral types were not published, an estimate of $\pm{1}$ spectral subtype was adopted. Where a range of possible spectral types was quoted from a single source for a particular star, the median spectral type of the published range was adopted. In this case, the error on the spectral type was adjusted to account for the increased range of possible values. For instance, a star with a published value of spectral type given as K2-K7 would be assigned a spectral type of K4.5 and an error of $\pm2$ spectral subtypes. 

\subsection{Bolometric luminosities}\label{Lbol_section}
\begin{figure}
 \centering
 \includegraphics[width=0.45\textwidth]{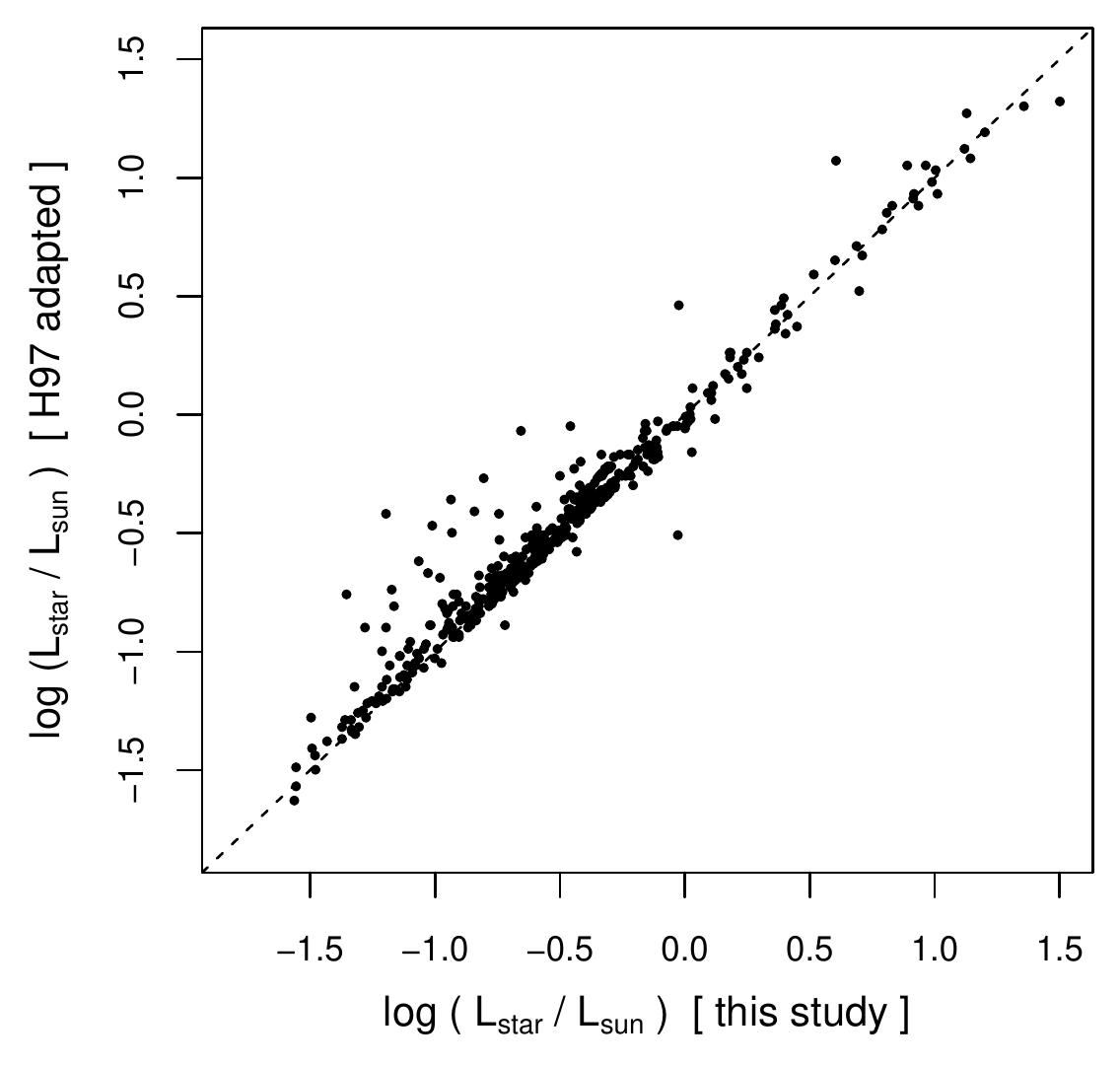}
 \centering
 \caption{Comparison between newly calculated luminosities for our ONC sample and those of \citet{Hillenbrand1997}, adjusted to account for the different distance modulus and solar bolometric luminosity used in this study. The dashed line shows a one-to-one fit to the data. In general, good agreement is found. The main source of spread is caused by the use of different spectral types.}
 \label{lum_comparison}
\end{figure}

\begin{figure*}
 \centering
 \includegraphics[width=7in]{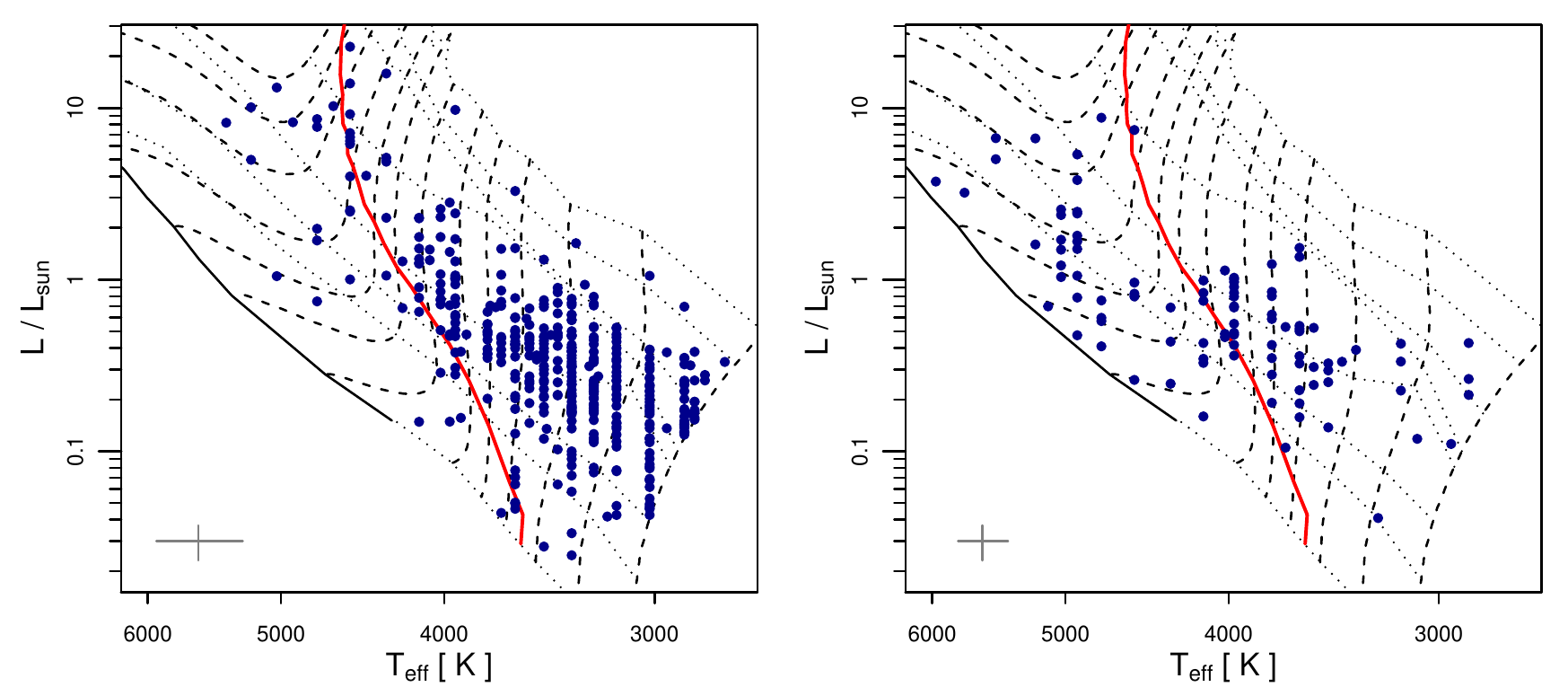}
 \centering
 \caption{HR diagrams constructed from the \citet{Siess2000} PMS evolutionary models for the ONC sample (left) and Taurus-Auriga (right). The mass tracks (dashed black lines) are shown (from right to left) for $0.1$, $0.2$, $0.3$, $0.4$, $0.5$, $0.6$, $0.8$, $1.0$, $1.2$, $1.5$, $2.0$, $2.5$, and $3.0\,\rmn{M_\odot}$ stars. Isochrones (black dotted lines) are shown (from upper right to lower left) for ages $0.01$, $0.05$, $0.2$, $0.5$, $2$, $5$, $10$, and $60\,$Myr. The position of the ZAMS is shown as a solid black line for $0.7$--$3.0\rmn{M_\odot}$ stars. The solid red lines marks the age at which each mass of star develops a radiative core according to equation (\ref{tcore}). Stars that fell to the left of this line were removed from further analysis. An average error bar is included for reference in the lower left of each plot.} 
 \label{hr_plot}
\end{figure*}

$L_{\star}$ can be calculated from the application of a bolometric correction to a single distance modulus- and extinction-corrected optical apparent magnitude \citep{Hillenbrand1997}. However, the choice of waveband is crucial in order to ensure only photospheric emission is observed. For Class II objects, $U$- and $B$-band magnitudes are unsuitable as they contain additional emission resulting from accretion. Similarly, $J$-, $H$-, and $K$-bands are unsuitable as they can contain excess emission from dust. We follow \citet{Hillenbrand1997} and use Cousins $I_\rmn{c}$-band photometry to calculate $L_{\star}$, ensuring that both accretion and disc emission remain minimal.

As in \citet{Hillenbrand1997}, we calculate luminosities from the observed photometry,
\begin{eqnarray}
	\log\left(\frac{L_{\rmn{\star}}}{L_{\odot}}\right) & = & 0.4 [M_{\rmn{bol},\odot}-(I_{\rmn{c}}-A_{\rmn{I_{c}}})+DM\\
	\nonumber & & - BC_{\rmn{I_{c}}}(T_{\rmn{eff}})].
\end{eqnarray} 
 Here, $M_{\rmn{bol},\odot}=4.755\,$mag is the bolometric absolute magnitude of the Sun \citep{Mamajek2012}\footnote{We consistently use physical constants and solar values from Eric Mamajek's ``Basic Astronomical Data for the Sun'' (http://sites.google.com/site/mamajeksstarnotes/basic-astronomical-data-for-the-sun) throughout this study.}, $I_{\rmn{c}}$ is the apparent magnitude of emission in the Cousins $I_\rmn{c}$-band, $A_{\rmn{I_{c}}}$ is the extinction at $I_{\rmn{c}}$, $DM$ is the distance modulus, and $BC_{\rmn{I_{c}}}(T_{\rmn{eff}})$ is the temperature-dependent bolometric correction at $I_{\rmn{c}}$. 

We used spectral type-dependent intrinsic colours, $(V-I_{\rmn{c}})_{0}$, and $V$-band bolometric corrections, $BC_{\rmn{V}}(T_{\rmn{eff}})$, presented in \citet[see Section \ref{SpT_Teff_section} for the reasoning behind the use of these models]{Pecaut2013}, to derive individual values of $BC_{\rmn{I_{c}}}(T_{\rmn{eff}})$ such that 
\begin{equation}
	BC_{\rmn{I_{c}}}(T_{\rmn{eff}})=BC_{\rmn{V}}(T_{\rmn{eff}})+(V-I_{\rmn{c}})_{0}.
\end{equation} 

We adopt the extinction law of \citet{Rieke1985}, transformed from the Johnsons to the Cousins photometric system by \citet{Hillenbrand1997},  
\begin{equation}
	A_{\rmn{I_{c}}}=0.61A_{\rmn{V}}=1.56[(V-I_{\rmn{c}})-(V-I_{\rmn{c}})_{0}],
\end{equation}
where $(V-I_{\rmn{c}})_{0}$ is the intrinsic colour appropriate to the spectral type of the star and $(V-I_{\rmn{c}})$ is the observed value. In the case where negative values of extinction were calculated, this indicated that the observed colours of that star were too blue for the assigned spectral type. For these stars, $A_{\rmn{I_{c}}}$ was set equal to zero. Consequently, the $L_{\star}$ calculated for these stars are lower limits and are considered as such in the following analysis. 

The distance modulus is assumed to be constant for all stars within each star forming region. For the ONC, we adopt a distance of $414\pm7\,$pc \citep{Menten2007} and, for Taurus-Auriga, we adopt $140\pm20\,$pc \citep{Elias1978, Loinard2007, Torres2009, Torres2012}. 

$V$- and $I_{\rmn{c}}$-band photometry for the ONC was taken from \citet{Hillenbrand1997}. Fig. \ref{lum_comparison} compares the new $L_{\star}$ for our ONC sample against the $L_{\star}$ from \citet{Hillenbrand1997}, adjusted to account for the different distance modulus and solar bolometric magnitude we have used. In the majority of cases, our updated $L_{\star}$ agree well with those in \citet{Hillenbrand1997}. The main source of spread can be attributed to our usage of updated spectral types. 

For the Taurus-Auriga region, individual sources of $V$- and $I_{\rmn{c}}$-band photometry are detailed in Table \ref{TAUtable}. Due to the periodic nature of the stars in our sample (Section \ref{Prot_section}), only data from studies that took contemporaneous measurements in both wavebands were included. The number of members of the Taurus-Auriga star forming region ($\sim348$ \citealt{Luhman2010}) is much smaller than that of the ONC ($>1000$ \citealt{Dario2010}) and the region has higher levels of optical extinction. These differences mean that the Taurus-Auriga sample is much smaller than the ONC sample. To attempt to counter this, we also obtained $B$- and $V$-band photometry which enabled us to calculate $L_{\star}$ from $V$-band magnitudes for an additional $23$ stars in Taurus-Auriga. In this case, 
\begin{eqnarray}
	\log\left(\frac{L_{\star}}{L_{\odot}}\right) & = & 0.4[M_{\rmn{bol},\odot}-(V-A_{\rmn{V}})+DM \\
	\nonumber & & - BC_{\rmn{V}}(T_{\rmn{eff}})],
\end{eqnarray}
 where the extinction at $V$ is taken from \citet{Rieke1985} such that
\begin{equation}
	A_{\rmn{V}}=3.09[(B-V)-(B-V)_{\rmn{0}}].
\end{equation}
 Here, $(B-V)_{\rmn{0}}$ is the intrinsic $(B-V)$ colour, again taken from \citet{Pecaut2013}.

Our choice of waveband should reduce the level of contamination by sources other than pure photospheric emission. However, as we make no attempt to calculate the accretion luminosity for any of the stars in our sample, our $L_{\star}$ may be underestimated for the most active accretors due to underestimated extinction values \citep{Hillenbrand1997, Dario2010}. Additionally, our method may lead to the underestimation of $L_{\star}$ for stars hosting dense discs at high inclinations \citep{Kraus2009}. We attempt to take both of these effects into account by assuming a conservative error estimate of $\pm0.1\,$dex in $\log\left(L_{\rmn{\star}}/{L_{\rmn{\odot}}}\right)$ for all stars. 

\subsubsection{Multiplicity}\label{multiplicity}
The presence of binaries and multiples can bias our study in various ways. For instance, a companion surrounded by an extended dusty disc or torus may be able to produce photometric variability on time-scales similar to stellar rotation periods \citep{Percy2010}. Alternatively, if the photometry used to calculate $L_{\star}$ includes a component from an unresolved companion, it can effect the placement of the star on the HR diagram \citep{Hartmann2001}, making the star appear systematically brighter and therefore younger. This effect is more problematic for regions of star formation older than $\sim15\,$Myrs \citep{Preibisch2012, Soderblom2013} as the spacing between the isochrones is smaller (e.g. Fig. \ref{hr_plot}; Section \ref{massage}), producing systematically overestimated luminosities. More problematic at the age of the ONC and Taurus-Auriga ($\sim1-2\,$Myrs; \citealt{Hillenbrand1997, White2001}) are the systematic errors on $L_{\star}$ associated with differential extinction and variable accretion \citep{Soderblom2013}, which we address in Section \ref{radiispread}. 

To minimise the effects of multiplicity, we cross-checked our ONC and Taurus-Auriga samples against previous studies of multiple stellar systems in both regions (\citealt{Leinert1993, Nordstrom1994, Mathieu1994, Osterloh1995, Duchene1999, Oh2006, Reipurth2007, Kraus2007, Furesz2008, Tobin2009, Luhman2009, Luhman2010, Rebull2010, Cieza2012}; \citealt*{Daemgen2012}; \citealt{Harris2012, Andrews2013, Correia2013}) and removed all those identified as binary or multiple systems. In addition, stars were also removed if their spectroscopy suggested the existence of an unresolved companion \citep{Morales2011, Hillenbrand2013}. 

\subsection{Stellar masses and ages}\label{massage}
Stellar masses and ages were calculated from $T_{\rmn{eff}}$ and $L_{\star}$ using \citet{Siess2000} PMS model isochrone fitting. The models are applicable to stars above $0.1\,\rmn{M_\odot}$ and we apply an upper mass limit of $3.0\,\rmn{M_\odot}$ as stars more massive than this are not T Tauri stars. The range of stellar ages covered by the models extends from the stellar birth line to the ZAMS but a star older than $\sim10\,\rmn{Myr}$ is unlikely to be a member of the ONC or Taurus-Auriga (see Section \ref{agespread}). The corresponding HR diagrams for our ONC and Taurus-Auriga samples are shown in Fig. \ref{hr_plot}. Stars that lay outside of the imposed boundaries could not be assigned a stellar mass or age. 

We use the \citet{Siess2000} PMS evolutionary models to translate the errors in $L_{\star}$ and $T_{\rmn{eff}}$ (in terms of the error in spectral type) into estimates of errors on stellar mass and age. We do not consider errors within the PMS evolutionary models themselves; a discussion of these can be found in \citet{Siess2001}. The errors on $\log\left(T_{\rmn{eff}}\right)$ and $\log\left(L_{\star}/{L_{\rmn{\odot}}}\right)$ define the major and minor axes of an ellipse in the HR diagram. The corresponding ellipse in $M_{\star}\,$--$\,Age$ space was calculated by iteratively tracing around the outside of the ellipse in $\log\left(T_{\rmn{eff}}\right)\,$--$\,\log\left(L_{\star}/{L_{\odot}}\right)$ space and calculating the stellar mass and age at each point. The maximum and minimum values of stellar mass and age calculated via this process were then used to estimate errors on the stellar mass and age for each star. 

This method results in upper and lower bounded errors that are not symmetric with the difference being most apparent for the errors on the age estimates. Contraction occurs on a Kelvin-Helmholtz time-scale, $t_{\rmn{KH}}\propto 1/R_{\star}$, such that the rate of contraction slows with time. This causes the isochrones to ``bunch up'' in the HR diagram at older ages, producing upper bounded age errors that exceed the lower bounded errors.

Where the errors in $\log\left(T_{\rmn{eff}}\right)\,$--$\,\log\left(L_{\star}/{L_{\odot}}\right)$ space exceeded the bounds imposed by the \citet{Siess2000} model limits, the upper and lower bounds to stellar masses and ages were assigned individually after conservative, by-eye inspection of the HR diagram. 

\subsection{Stellar radii}\label{radiispread}
Under the assumption that the spread in $L_{\star}$ observed in our ONC and Taurus-Auriga samples is indicative of a real spread in stellar radii, we calculate $R_{\star}$ directly from $T_{\rmn{eff}}$ and $L_{\star}$ using
\begin{equation}
	\frac{R_{\rmn{\star}}}{R_{\rmn{\odot}}}=\left(\frac{L_{\rmn{\star}}}{L_{\rmn{\odot}}}\right)^{1/2}\left(\frac{T_{\rmn{eff,\odot}}}{T_{\rmn{eff,\star}}}\right)^{2},
\end{equation}
where $T_{\rmn{eff,\odot}}=5771.8\,$K \citep{Mamajek2012b}. 

It has been suggested that the observed spread in $L_{\star}$ is a consequence of a combination of observational and astrophysical uncertainties such as contamination by unresolved binaries, photometric variability, and inadequate correction for variable extinction \citep{Hartmann2001} rather than of a true spread in $R_{\star}$. Although these effects all contribute to the differences in $L_{\star}$ throughout the regions considered, they have been found not to explain the full scale of the observed spreads (\citealt{Burningham2005, Preibisch2005, Dario2010}; \citealt*{Hillenbrand2008}; \citealt*{Slesnick2008}; \citealt{Preibisch2012}). In addition, \citet{Jeffries2007b} estimated $R_{\star}$ independently of $L_{\star}$ and $T_{\rmn{eff}}$ by combining projected stellar rotational velocities and rotational periods for stars in the ONC. Even after accounting for observational uncertainties and random inclinations of the stellar rotation axes, the spread in $R_{\star}$ was still observed.

\subsubsection{Luminosity spreads as indicators of true age spreads}\label{agespread}
Our use of $T_{\rmn{eff}}$ and $L_{\star}$ to derive individual ages for the stars in our sample (Section \ref{massage}) further assumes that the observed spread in $L_{\star}$ (which we have attributed to a real spread in $R_{\star}$) corresponds to a real spread in age. The question of whether this assumption is correct has been heavily debated in the literature (see e.g. \citealt{Jeffries2012} and \citealt{Soderblom2013} for recent reviews).

Certain studies have argued that the observed spread in $R_{\star}$ is produced by magnetic effects reducing convective efficiency or significant spot coverage on the stellar surface (e.g. \citealt{Spruit1986, Jackson2014}). However, \citet*{Chabrier2007}, \citet{Morales2010} and \citet{Feiden2014} find that the level of radius inflation produced by the inhibition of convective efficiency is negligible ($\sim0.1-2\%$) for fully convective stars. In addition, by considering observed spot temperatures of K and early-M stars at $82-90\%$ of photospheric temperature \citep{Boyajian2012} and observed spot coverage of a few percent to $\sim40\%$ (\citealt*{ONeal1998}; \citealt{Barnes2001, Barnes2004, ONeal2004, Morin2008a, Hackman2012}), \citet{Feiden2014} found that starspots could only produce the degree of radial inflation inferred from $L_{\star}$ spreads if unattainably high interior magnetic field strengths were present.

Alternatively, episodic accretion during the assembly phase with mass accretion rates $\geq10^{-5}\,\rmn{M_{\odot}}\,\rmn{yr^{-1}}$ has been proposed as a method of producing the observed spread in stellar radii (\citealt*{Tout1999}; \citealt{Baraffe2002}; \citealt*{Baraffe2009}). Depending on the amount of accretion kinetic energy absorbed by the star during this phase, the star can either contract at a greater rate and then remain at almost constant radius for $\sim10\,$Myrs or it can inflate to larger radii before quickly contracting back to the non-accreting isochrone expected of its mass and age \citep{Baraffe2009, Littlefair2011}. Thus, the stellar radius would be more an indication of accretion history rather than age. However, the ability of this mechanism to produce the observed $L_{\star}$ spreads at low masses has been contested \citep{Hosokawa2011} and depends on the initial protostellar mass assumed in the ``cold accretion'' models \citep*{Baraffe2012}.

Using alternative age diagnostics such as lithium depletion levels has revealed that a few percent of ONC and Taurus-Auriga PMS members are consistent with being $\geq10\,$Myr old \citep{Palla2007, Sacco2007}. Furthermore, \citet{Sergison2013} determined the ages of stars within the ONC and NGC 2264 using lithium depletion and PMS isochrones, finding a modest correlation between the two age indicators. With this in mind, we assume that the age spreads in the ONC and Taurus-Auriga are real and we use the individual isochronal ages to study the evolution of $j_{\star}$.

\subsection{Rotation periods}\label{Prot_section}
\begin{figure}
 \centering
 \includegraphics[width=0.45\textwidth]{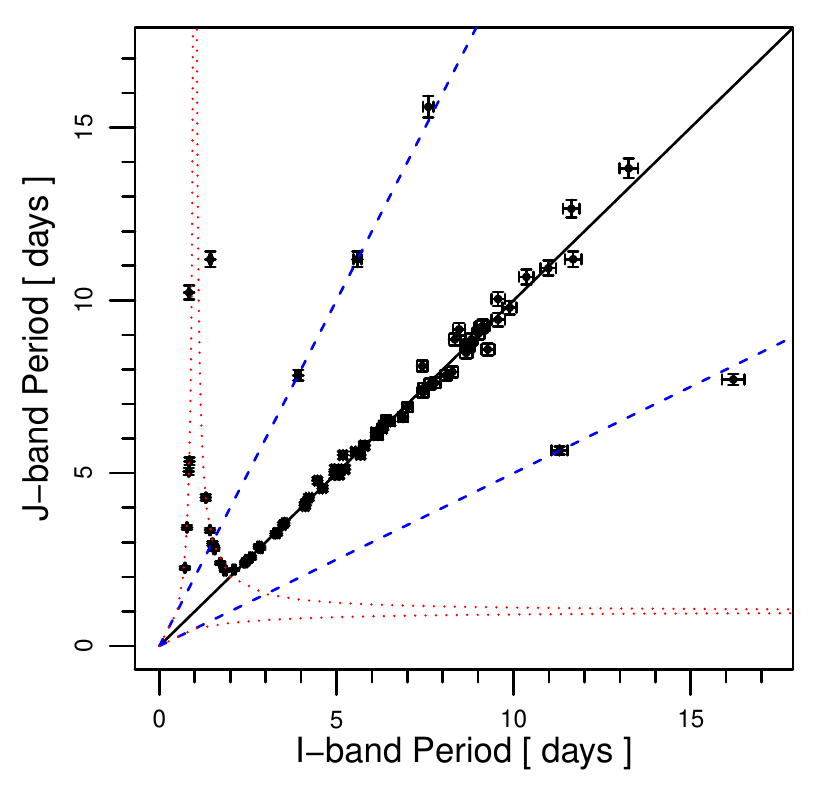}
 \centering
 \caption{Example of the analysis undertaken to check for the effects of beats (red dotted lines) and harmonics (blue dashed lines) in the ONC sample. The $I$-band periods are from \citet{Herbst2002, Parihar2009} and \citet*{Rodriguez2009} and the $J$-band periods are from \citep*{Carpenter2001}. Stars located on the solid black line have the same measured rotation period in both the optical and NIR.}
 \label{optIRperiods}
\end{figure}

The periodic nature of PMS stars has been used to determine stellar rotation periods using both optical and infra-red (IR) wavelengths. For both Class II and Class III PMS stars, this observed periodicity can be attributed to cool starspots on the stellar surface. These reduce the flux received from the star at a rate determined by its rotational period \citep{Carpenter2001, DeWarf2003, Grankin2008, Frasca2009}. Additionally, for Class II PMS stars, magnetospheric accretion of disc material can produce hotspots on the stellar surface. These hotspots lead to an increase in flux received from the star, modulated by rotation in the same way as for the cool starspots. 

Periodic flux changes in the near-IR (NIR) and mid-IR (MIR) can also be caused by temperate, opacity, or geometric changes in the inner disc (\citealt{Bouvier2003, Alencar2010, Morales2011}; \citealt*{Artemenko2012}; \citealt{Cody2014}). When these changes arise from regions in the disc close to the corotation radius, they can be used as indicators of stellar surface rotation rates. 

Problems with the measurement of stellar rotation periods arise if multiple sources of periodicity are present. In such a case, the measured rotation period may only be a fraction of its actual value. Additionally, if observations are taken at a single longitude, the Earth's day-night cycle imposes a one day sampling interval such that rotation periods of $\sim1\,$day can have a beat period, $B$, recorded rather than the true rotational period, $P$, \citep{Cieza2006} where
\begin{equation}\label{beats}
	\frac{1}{B}=\pm1\,\rmn{day^{-1}}\,\pm\,\frac{1}{P}.
\end{equation}

We gathered previously published rotation periods from the literature as detailed in Tables \ref{ONCtable} and \ref{TAUtable} for the ONC and Taurus-Auriga samples, respectively. Where errors for the rotation period were not reported, a conservative estimate of $0.01\,$days was assumed. In the cases where multiple rotation periods were available for the same star, we checked for the effects of harmonics and beats, described above. An example of this analysis is shown in Fig. \ref{optIRperiods} for stars in the ONC. Any rotation periods that appeared to show evidence of these phenomena were removed from Tables \ref{ONCtable} and \ref{TAUtable}. 

All rotation periods for members of Taurus-Auriga were measured at optical wavebands whereas those for members of the ONC were measured at optical, NIR, or MIR wavebands. A general agreement between rotation periods measured at optical and NIR wavelengths was found, as shown in Fig.~\ref{optIRperiods}. The major differences between the optical and IR rotation periods can be explained in terms of either harmonics and beats phenomena. We flagged all rotation periods longer than $15\,$days and removed them from further analysis. It is unlikely that these trace photospheric rotation and are more likely to be caused by occultation of the stellar surface by disc material exterior to the co-rotation radius (\citealt*{Artemenko2010}; \citealt{Cody2014}). 

\subsubsection{Central densities and the radius of gyration}
\begin{figure}
 \centering
 \includegraphics[width=0.45\textwidth]{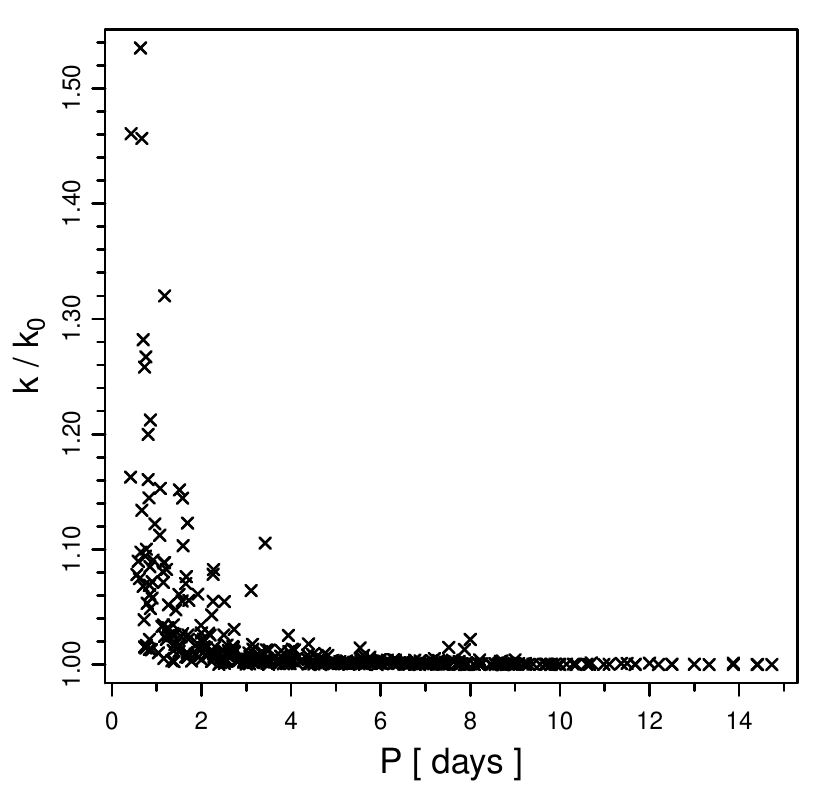}
 \centering
 \caption{The relationship between stellar rotation period, $P$, and the radius of gyration, $k$, normalised to that of a perfect sphere, $k_{0}=(2/3)^{1/2}$ for the fully convective stars in the ONC and Taurus-Auriga with $P<15\,$days. For all but a few of the fastest rotators, the stars in our sample are well approximated by perfect spheres.} 
 \label{gyration}
\end{figure}

Once the ages and stellar masses had been calculated and the fully convective limit imposed (see Section \ref{massage}), we linearly interpolated the individual central densities from \citet{Siess2000} PMS core isochrones. Combining these with the stellar rotation periods allowed us to calculate the radius of gyration, $k$, for each individual source using equation (\ref{gyration}). Fig. \ref{gyration} shows the relationship between the radius of gyration and the rotation period for the fully convective stars in our combined ONC and Taurus-Auriga sample with $P<15\,$days. For all but a few of the fastest rotators, the stars are well approximated by perfect spheres.

As a final check of the rotation periods, we ensured that none of the rapidly rotating stars in Fig. \ref{gyration} appeared to be rotating at rates exceeding break-up velocity. An object of mass, $M_{\star}$, rotation period, $P$, and equatorial radius, $R_{\rmn{eq}}$, will break apart if the acceleration due to the centripetal force, 
\begin{equation}
	a_{\rmn{cent}}=\frac{4\pi^{2} R_{\rmn{eq}}}{P^{2}},
\end{equation}
 exceeds the acceleration due to gravity, 
\begin{equation}
	a_{\rmn{grav}}=\frac{GM_{\star}}{R_{\rmn{eq}}^{2}}. 
\end{equation}
For the more rapidly rotating stars, the stellar shape will differ from a perfect sphere with the star becoming more distended around its equatorial regions. Following \citet{Chandrasekhar1935},
\begin{equation}
	\frac{R_{\rmn{eq}}}{R_{\star \rmn{,0}}}=a-bP_{2}\left(\theta=0\right) 
\end{equation}
 where $a$ and $b$ are given by equations (\ref{a}) and (\ref{b}), $R_{\star \rmn{,0}}$ is the radius of a non-rotating star, $\theta$ is the usual polar angle, and $P_{2}(\theta=0)$ is the second order Legendre polynomial. This enables us to define a critical rotation period, 
\begin{equation}
	P_{\rmn{crit}}=\frac{2\pi R_{\rmn{eq}}^{3/2}}{\left(GM_{\star}\right)^{1/2}},
\end{equation} 
such that if $P<P_{\rmn{crit}}$, the object will break apart. 

We can use the critical rotation period to define a critical specific stellar AM, $j_{\rmn{crit}}$, using equation (\ref{Jstar}). We remove any stars from our sample for which
\begin{equation} 
	\frac{j_{\star}}{j_{\rmn{crit}}}=\frac{2\pi R_{\rmn{eq}}^{3/2}}{\left(GM_{\star}\right)^{1/2}P}>1.
\end{equation}

We found ten of the fully convective stars in our ONC sample have $j_{\star}\geq j_{\rmn{crit}}$. All of these stars are at the low mass end of our sample, with spectral types of M3.5 to M5.5. Nine of the ten fully convective stars for which $j_{\star}\geq j_{\rmn{crit}}$ have only one recorded rotation period, each measured at a fraction of a day. It is possible that these rotation periods are affected by the beat phenomena. For the other fully convective star, the rotation period is measured at $1.18\,$days but its luminosity is very high for its spectral type (M5). Thus, the stellar radius is much larger than other stars of a similar spectral type. It is possible that this luminosity is overestimated for this star, perhaps due to the presence of an unresolved binary (see Section \ref{multiplicity}). 

\subsubsection{Cluster membership: removing contamination from period distributions}\label{membership}
The ONC is part of the much larger Orion A cloud and is surrounded by neighbouring regions of star formation \citep{Hillenbrand1997, Alves2012, Bouy2014}. In previous studies of ONC rotation period distributions, the inclusion of these regions has blurred the location of the peak of rapid rotators as well as the respective height of the two peaks, producing a unimodal distribution (e.g. \citealt{Rebull2001}). For this reason, it was imperative to ensure that our ONC sample was as clean as possible. Members of the Orion flanking fields and other neighbouring regions such as NGC 1980, L1641N, L1641W, and NGC 1981 were removed from the sample. Only stars that lay within the ``traditional" region of the ONC ($84.1^{\circ}\leq$ RA $\leq83.0^{\circ}$ and $-5.0^{\circ}\leq$ dec $\leq-5.7^{\circ}$; \citealt{CiezaBaliber2007}) were retained. 

Even with this cut applied to right ascension and declination, the ONC sample could be contaminated by members of the somewhat overlapping clusters L1641N and NGC 1980 \citep{Alves2012, Bouy2014}. All objects listed in \citet{Hillenbrand1997} as having a membership probability $<98\%$ were removed from the sample and any additional non-members were removed by cross-referencing with the recent studies of \citet{Fang2013} and \citet{Pillitteri2013}. 

\subsubsection{Mass segregation}\label{masscut}
Due to the presence of a mass-rotation relation in PMS stars \citep{Herbst2000, CiezaBaliber2007}, we split our sample into two, mass-segregated groups. We do not base our mass cut directly on stellar mass due to the apparent age spread in our samples (Fig. \ref{hr_plot}). Instead, we base our mass cut on spectral type with the ``high mass'' sample having a spectral type of M2 or earlier and the ``low mass'' stars being later than M2. This choice of spectral type cut off is based on the work of \citet{CiezaBaliber2007} who looked at the effect of varying the location of the spectral type cut on period distributions in the ONC. They found their results were consistent with a sudden change in stellar magnetic field strength or structure between the M2 and M3 spectral types and that the bimodality of the ``high mass'' sample was severely affected by even a small contamination by lower mass stars \citep{CiezaBaliber2007}. At the range of ages included in our sample, this spectral type corresponds to a stellar mass of $\sim0.35\,\rmn{M_{\odot}}$, the mass below which stars remain fully convective during their MS lifetimes (see Section \ref{AMtheory}). 

\subsection{Disc diagnostics}\label{disc_diagnosis}
Observations of an accretion disc-rotation relation are dependent on the use of a reliable method to identify accretion discs. In the absence of circumstellar dust, the spectral energy distribution (SED) of a PMS star is purely photospheric in origin and resembles that of a black-body \citep{Lada1984, Lada1987}. During this phase, the PMS star is referred to as a Class III object. At earlier stages of formation, circumstellar dust is present around the star. This dust reprocesses incident stellar emission at longer wavelengths, giving rise to excess emission in the IR part of the SED. When this material fully envelopes the star, it is referred to as a Class I object. The rotation of the enveloping material around the central protostar means that high latitude regions of the envelope have lower angular momentum than lower latitudes. Consequently, the infalling material flattens into a disc, allowing the stellar photosphere to become optically visible, and the star is identified as a Class II object \citep*{Adams1987}. 

For most of the stars in our Taurus-Auriga sample, the results of detailed SED modelling are available in the literature. We used these to identify the presence (or absence) of an IR excess and so define source as a Class II (or Class III) object \citep{Kenyon1995, Andrews2005, Luhman2010, Rebull2010}. In addition, resolved (sub)millimetre observations have directly imaged discs around individual stars in Taurus-Auriga \citep{Kitamura2002, Rodmann2006, Andrews2007, Guilloteau2011}. We used these identifications to supplement our Taurus-Auriga Class II sample. 

For the ONC sample, the same detailed SED modelling was not available. Most early studies of PMS AM evolution in the ONC relied on NIR excesses and $\rmn{H}\alpha$ equivalent widths (EW) to ascertain whether a PMS star hosted a circumstellar disc or was actively accreting, respectively. However, the magnitude of the IR excess at NIR wavelengths, and the $\rmn{H}\alpha$ EW, are dependent on stellar mass \citep{White2003, Littlefair2005, Cody2014} and, due to the low contrast between photospheric and disc emission at NIR wavelengths, disc indicators relying on $J$-, $H$-, or $K$-band emission were found to miss up to $30\%$ of discs detected at longer wavelengths \citep{Hillenbrand1998}. In addition, observations used to derive NIR excesses were often taken at different epochs and were consequently affected by the intrinsically periodic nature of PMS stars (Section \ref{Prot_section}). More recently, $\itl{Spitzer}$ IRAC have provided high resolution, contemporaneous observations between $3.6$ and $70\,\rmn{\mu m}$ which allow for more reliable PMS classification. 

We gathered $2.2$, $3.6$, $4.5$, $5.8$, and $8.0\,\rmn{\mu m}$ $\itl{Spitzer}$ IRAC fluxes from \citet{Rebull2006}, \citet{CiezaBaliber2007}, \citet{Prisinzano2008}, and \citet{Morales2011}. Over this wavelength range, the spectral index, $\alpha_{\rmn{i-\left(i+1\right)}}$, \citep{Lada1987} is defined as
\begin{equation}
	\alpha_{\rmn{i-\left(i+1\right)}}=-\frac{\log\left(\lambda_{\rmn{i+1}}F_{\rmn{\lambda_{i+1}}}\right)-\log\left(\lambda_{\rmn{i}}F_{\rmn{\lambda_{i}}}\right)}{\log\left(\lambda_{\rmn{i+1}}\right)-\log\left(\lambda_{\rmn{i}}\right)},
\end{equation}
where $i = 1, 2, 3, 4$ and refers to the waveband such that $\left[2.2, 3.6, 4.5, 5.8\right]\,\rmn{\mu m}$ are wavelengths $\left[\lambda_1, \lambda_2, \lambda_3, \lambda_4\right]$.  

We made sure that our Class II and Class III samples were not contaminated by more embedded objects. Any source that displayed an increasing SED over the $2.2$--$8.0\,\rmn{\mu m}$ wavelength range was removed from further analysis. We made no attempt to classify these objects as either Class 0 or Class I and these objects are included in Tables \ref{ONCtable} and \ref{TAUtable} amongst the unclassified sources. 

To identify the Class II sources in our ONC sample, we followed methods employed in \citet{Hartmann2005} and \citet{Rebull2006}. We selected all sources for which $\left[3.6\right]-\left[8.0\right]>1.0$, or $0.2<\left[3.6\right]-\left[4.5\right]<0.7$ and $0.6<\left[5.8\right]-\left[8.0\right]<1.0$. These are slightly more restrictive criteria than others employed using $\itl{Spitzer}$ IRAC colours (e.g. \citealt{Megeath2004}) but should enable us to compile as pure a set of Class II objects as possible. Identification as Class II required agreement between the four studies from which we took the $\itl{Spitzer}$ IRAC fluxes. Where identifications did not agree, the source remained unclassified. It is hoped that this will reduce contamination from transitional discs and ``flat'' spectrum objects. 

The Class III objects were selected from the remaining unclassified sources. A Class III PMS star displays purely photospheric emission as it lacks the IR excess seen for disced objects. As such, to be identified as a Class III object, a star must satisfy $\left[2.2\right]-\left[3.6\right]<0.5$, $\left[3.6\right]-\left[4.5\right]<0.2$, $\left[4.5\right]-\left[5.8\right]<0.2$, and $\left[5.8\right]-\left[8.0\right]<0.2$ \citep{Prisinzano2008}. Alternatively, objects were also identified as purely photospheric if they were detected at $I_{\rmn{c}}$ band but not detected at wavelengths longer than $3.6\,\rmn{\mu m}$. Again, just as with the Class II sample, agreement between the sources of $\itl{Spitzer}$ IRAC fluxes was required in order for the source to be identified as Class III.

\begin{landscape}
\begin{table}
\begin{minipage}{235mm}
\centering
 \caption{Stellar data for all members of the ONC not identified as binary or multiple for which a spectral type was available in the literature (see Section \ref{membership} for membership constraints). The full table is available in electronic form in the Supplementary Materials section. A sample is given here to illustrate its content. Column 1 gives the SIMBAD identification for the star; columns 2, 3, and 4 list the adopted spectral type (SpT), its error in spectral subtype, and the reference as in \citet{Hillenbrand2013} (H13; except [D10] -- \citealt{Dario2010}); columns 5 and 6 list the effective temperature, $T_{\rmn{eff}}$, and logarithmic bolometric luminosity, $\log(L_{\star}/L_{\odot})$, calculated from SpT and optical photometry (see Sections \ref{SpT_Teff_section} and \ref{Lbol_section} for details); columns 7 and 8 list the stellar mass in solar units and the age in Myr, respectively, calculated from $T_{\rmn{eff}}$ and $\log(L_{\star}/L_{\odot})$ using \citet{Siess2000} PMS evolutionary models (see Section \ref{massage}); column 9 lists the stellar radius in solar units; columns 10, 11, and 12 list the adopted rotation period in days, the observed waveband for rotation period measurement (opt -- optical, NIR -- near infra-red, MIR -- mid infra-red), and the reference for the rotation period (E93 -- \citealt{Edwards1993}, G95 -- \citealt*{Gagne1995}, C96 -- \citealt{Choi1996}, S99 -- \citealt{Stassun1999}, H00 -- \citealt{Herbst2000}, C01 -- \citealt{Carpenter2001}, Re01 -- \citealt{Rebull2001}, Rh01 -- \citealt*{Rhode2001}, H02 -- \citealt{Herbst2002}, F09 -- \citealt{Frasca2009}, P09 -- \citealt{Parihar2009}, R09 -- \citealt{Rodriguez2009}, M11 -- \citealt{Morales2011}); columns 13 and 14 list the source classification based on MIR excess measurements (II -- disced, III -- disc-less), and based on the $\rmn{EW\left(CaII\right)}$ (A -- accreting, N -- not accreting) as detailed in Sections \ref{disc_diagnosis} and \ref{am_evol_section}, respectively; column 15 lists the reason, where applicable, for a star's exclusion from the final analysis (a -- not fully convective (see equation \ref{tcore} and Section \ref{AMtheory}), b -- star lies outside the limits imposed in isochronal fitting (see Section \ref{massage}), c -- no optical photometry available to calculate $L_{\star}$ (see Section \ref{Lbol_section}), d -- $P>15\,$days (see Section \ref{Prot_section}), e -- $j_{\star}>j_{\rmn{crit}}$ (see Section \ref{Prot_section}), f -- no reliable rotation period available, g -- able to calculate $j_{\star}$ but not able to classify source as II or III).}
 \begin{tabular}{@{}lccccccccccccccc@{}} 
 \hline
SIMBAD & SpT & $\sigma\left(\rmn{SpT}\right)$ & Ref & $T_{\rmn{eff}}$ & $\log\left(L_{\star}/L_{\odot}\right)$ & $M_{\star}$ & Age & $R_{\star}$ & Period & Obs & Ref & Class & Accretion & Notes\\ 
 & & & & $\left(\rmn{K}\right)$ &  & $\left(\rmn{M_{\odot}}\right)$ & $\left(\rmn{Myr}\right)$ & $\left(\rmn{R_{\odot}}\right)$ & $\left(\rmn{days}\right)$ & & & & & \\
(1) & (2) & (3) & (4) & (5) & (6) & (7) & (8) & (9) & (10) & (11) & (12) & (13) & (14) & (15)\\
 \hline
LT Ori & G8 & $1$ & Ste & $5210$ & $1.128$ & $2.73_{-0.21}^{+0.27}$ & $1.80_{-0.50}^{+0.60}$ & $4.49 \pm 0.54$ & $0.50$ & opt & G95 & --- & --- & a,e \\ 
V1229 Ori & K0 & $1$ & H & $5030$ & $1.119$ & $2.86_{-0.18}^{+0.13}$ & $1.20_{-0.30}^{+0.50}$ & $4.77 \pm 0.58$ & $14.30$ & opt & H00 & III & --- & a \\ 
V1963 Ori & G8 & $1$ & H & $5210$ & $1.004$ & $2.53_{-0.28}^{+0.22}$ & $2.20_{-0.60}^{+0.90}$ & $3.90 \pm 0.47$ & $3.37$ & MIR & M11 & III & N & a \\ 
V2235 Ori & K1 & $1$ & H & $4920$ & $0.917$ & $2.47_{-0.27}^{+0.20}$ & $1.40_{-0.50}^{+0.80}$ & $3.96 \pm 0.52$ & $17.91$ & opt & H02 & II & A & a,d \\ 
V403 Ori & K3 & $1$ & H & $4550$ & $1.357$ & $2.51_{-1.02}^{+0.49}$ & $0.30_{-0.10}^{+0.10}$ & $7.68 \pm 1.15$ & $6.09$ & MIR & M11 & III & N & a \\ 
AK Ori & K2 & $1$ & Ste & $4760$ & $0.89$ & $2.21_{-0.49}^{+0.34}$ & $1.00_{-0.40}^{+0.80}$ & $4.09 \pm 0.59$ & $10.33$ & opt & G95 & II & --- & a \\ 
V1232 Ori & G6 & $1$ & H & $5390$ & $0.914$ & $2.20_{-0.25}^{+0.14}$ & $3.70_{-0.90}^{+1.50}$ & $3.28 \pm 0.40$ & $1.55$ & opt & H00 & III & N & a \\ 
V1509 Ori & K2.5 & $1$ & H & $4655$ & $1.011$ & $2.11_{-0.59}^{+0.52}$ & $0.60_{-0.20}^{+0.40}$ & $4.92 \pm 0.71$ & $6.99$ & opt & R09 & III & N & a \\ 
V426 Ori & K2 & $1$ & H & $4760$ & $0.935$ & $2.26_{-0.57}^{+0.34}$ & $0.90_{-0.40}^{+0.70}$ & $4.31 \pm 0.62$ & $5.15$ & opt & H02 & II & N & a \\ 
AF Ori & G8 & $3$ & H13 & $5210$ & $0.698$ & $2.00_{-0.30}^{+0.23}$ & $4.20_{-2.00}^{+3.40}$ & $2.74 \pm 0.44$ & --- & --- & --- & II & A & a,f \\ 
KM Ori & K3 & $1$ & Ste & $4550$ & $1.143$ & $2.08_{-0.85}^{+0.57}$ & $0.40_{-0.20}^{+0.20}$ & $6.00 \pm 0.90$ & $17.40$ & opt & H00 & III & --- & d \\ 
V348 Ori & K3 & $1$ & Ste & $4550$ & $0.964$ & $1.76_{-0.59}^{+0.63}$ & $0.50_{-0.20}^{+0.40}$ & $4.88 \pm 0.73$ & $8.71$ & opt & H00 & II & N & --- \\ 
V1331 Ori & K3e & $1$ & Sta & $4550$ & $0.854$ & $1.65_{-0.49}^{+0.58}$ & $0.60_{-0.20}^{+0.50}$ & $4.30 \pm 0.65$ & $10.70$ & opt & H00 & --- & N & g \\ 
V1444 Ori & K3 & $1$ & Ste & $4550$ & $0.79$ & $1.63_{-0.51}^{+0.49}$ & $0.70_{-0.30}^{+0.60}$ & $4.00 \pm 0.60$ & $3.45$ & opt & H00 & III & --- & --- \\ 
V2299 Ori & K3 & $1$ & Ste & $4550$ & $0.808$ & $1.66_{-0.51}^{+0.48}$ & $0.70_{-0.30}^{+0.60}$ & $4.08 \pm 0.61$ & --- & --- & --- & II & N & f \\ 
V1294 Ori & K3 & $1$ & Ste & $4550$ & $0.83$ & $1.60_{-0.41}^{+0.57}$ & $0.60_{-0.20}^{+0.60}$ & $4.18 \pm 0.63$ & $6.76$ & opt & H00 & III & --- & --- \\ 
V1333 Ori & K3 & $1$ & Ste & $4550$ & $0.601$ & $1.54_{-0.45}^{+0.40}$ & $1.10_{-0.50}^{+1.10}$ & $3.21 \pm 0.48$ & $9.23$ & opt & H00 & --- & N & a \\ 
V2140 Ori & K2 & $1$ & H & $4760$ & $0.296$ & $1.58_{-0.12}^{+0.12}$ & $4.30_{-1.90}^{+3.20}$ & $2.06 \pm 0.30$ & $3.82$ & opt & H02 & --- & N & a \\ 
V401 Ori & K2 & $1$ & H & $4760$ & $0.228$ & $1.51_{-0.13}^{+0.11}$ & $5.50_{-2.60}^{+3.70}$ & $1.91 \pm 0.28$ & $6.63$ & opt & S99 & II & --- & a \\ 
V356 Ori & K3 & $1$ & Ste & $4550$ & $0.403$ & $1.44_{-0.37}^{+0.29}$ & $1.80_{-0.80}^{+2.00}$ & $2.56 \pm 0.38$ & $1.57$ & opt & H00 & --- & N & a \\ 
V494 Ori & K3 & $1$ & H & $4550$ & $0.396$ & $1.46_{-0.40}^{+0.27}$ & $1.90_{-0.90}^{+1.90}$ & $2.54 \pm 0.38$ & --- & --- & --- & II & --- & a,f \\ 
AC Ori & K3.5 & $3$ & LR & $4450.5$ & $0.605$ & $1.30_{-0.69}^{+0.76}$ & $0.80_{-0.40}^{+2.60}$ & $3.38 \pm 0.88$ & --- & --- & --- & --- & A & f \\ 
MU Ori & K3 & $1$ & Ste & $4550$ & $0.001$ & $1.28_{-0.17}^{+0.09}$ & $6.90_{-3.60}^{+5.10}$ & $1.61 \pm 0.24$ & $2.22$ & opt & H02 & III & N & a \\ 
AE Ori & K4 & $1$ & Ste & $4330$ & $1.201$ & $1.26_{-0.20}^{+0.91}$ & $0.20_{-0.17}^{+0.10}$ & $7.10 \pm 1.09$ & $3.42$ & opt & H00 & III & N & --- \\ 
V1330 Ori & K4 & $1$ & H & $4330$ & $0.71$ & $1.15_{-0.35}^{+0.40}$ & $0.50_{-0.10}^{+0.30}$ & $4.03 \pm 0.62$ & $8.67$ & opt & H00 & III & --- & --- \\ 
V1337 Ori & K0 & $2$ & H & $5030$ & $0.02$ & $1.14_{-0.11}^{+0.19}$ & $17.50_{-8.00}^{+9.50}$ & $1.34 \pm 0.21$ & --- & --- & --- & II & N & a,f \\ 
LU Ori & K4 & $1$ & Ste & $4330$ & $0.687$ & $1.11_{-0.33}^{+0.48}$ & $0.50_{-0.10}^{+0.40}$ & $3.93 \pm 0.60$ & $4.08$ & opt & H00 & III & --- & --- \\ 
V1397 Ori & K2 & $1$ & H & $4760$ & $-0.127$ & $1.11_{-0.13}^{+0.13}$ & $16.00_{-6.60}^{+10.00}$ & $1.27 \pm 0.18$ & $5.41$ & opt & H00 & --- & --- & a \\ 
V377 Ori & K4 & $1$ & Ste & $4330$ & $0.36$ & $1.11_{-0.32}^{+0.36}$ & $1.20_{-0.50}^{+1.10}$ & $2.70 \pm 0.41$ & $13.00$ & opt & H02 & III & --- & --- \\ 
 \hline
\end{tabular} 
\label{ONCtable}
\end{minipage}
\end{table}
\end{landscape}

\begin{landscape}
\begin{table}
\begin{minipage}{235mm}
\centering
  \caption{Stellar data for all members of Taurus-Auriga not identified as binary or multiple for which a spectral type was available in the literature. The full table is available in electronic form in the Supplementary Materials section. A sample is given here to illustrate its content. Column 1 gives the SIMBAD identification for the star; columns 2 and 3 list the adopted spectral type (SpT) and its error in spectral subtype; column 4 lists the effective temperature, $T_{\rmn{eff}}$, calculated from SpT (see Section \ref{SpT_Teff_section} for details); columns 5 and 6 list the observed $V$-band magnitude and ($B$--$V$) colour; columns 7 and 8 list the observed $I_{\rmn{c}}$ magnitude and ($V$--$I_{\rmn{c}}$) colour; column 9 lists the adopted logarithmic bolometric luminosity, $\log(L_{\star}/L_{\odot})$ (see Section \ref{Lbol_section} for details); columns 10 and 11 list the stellar mass in solar units and the age in Myr, estimated from $T_{\rmn{eff}}$ and $\log(L_{\star}/L_{\odot})$ using \citet{Siess2000} PMS evolutionary models (see Section \ref{massage} for details); column 12 lists the stellar radius in solar units; column 13 lists the rotation period in days; column 14 lists the classification of the object based on SED fitting (II -- disced, III -- disc-less; see Section \ref{disc_diagnosis} for details); column 15 lists the references for the SpT, photometry, rotation period, and source classification ([1] -- \citealt{Cohen1979}, [2] -- \citealt{Bouvier1986}, [3] -- \citealt{Herbst1988}, [4] -- \citealt{Beckwith1990}, [5] -- \citealt{Bouvier1990}, [6] -- \citealt{Bouvier1993}, [7] -- \citealt{Edwards1993}, [8] -- \citealt{Grankin1993}, [9] -- \citealt{Herbst1994}, [10] -- \citealt{StromStrom1994}, [11] -- \citealt{Kenyon1995}, [12] -- \citealt{Fernandez1996}, [13] -- \citealt{Grankin1996}, [14] -- \citealt*{Osterloh1996}, [15] -- \citealt{Wichmann1996}, [16] -- \citealt{Bouvier1997}, [17] -- \citealt{Grankin1997}, [18] -- \citealt{Briceno1998}, [19] -- \citealt{Luhman1998}, [20] -- \citealt{Briceno1999}, [21] -- \citealt{Wichmann2000}, [22] -- \citealt{Mora2001}, [23] -- \citealt{Roberge2001}, [24] -- \citealt{Stassun2001}, [25] -- \citealt{White2001}, [26] -- \citealt{Briceno2002}, [27] -- \citealt{Vieira2003}, [28] -- \citealt{Luhman2004}, [29] -- \citealt{Andrews2005}, [30] -- \citealt{Massarotti2005}, [31] -- \citealt{Broeg2006}, [32] -- \citealt{Kundurthy2006}, [33] -- \citealt{Padgett2006}, [34] -- \citealt*{Scholz2006}, [35] -- \citealt*{Xing2006}, [36] -- \citealt{Grosso2007}, [37] -- \citealt{Chapillon2008}, [38] -- \citealt{Grankin2008}, [39] -- \citealt{Luhman2009}, [40] -- \citealt{Espaillat2010}, [41] -- \citealt{Luhman2010}, [42] -- \citealt{Rebull2010}, [43] -- \citealt{Andrews2011}, [44] -- \citealt{Furlan2011}, [45] -- \citealt{Xiao2012}, [46] -- \citealt{Cody2013}, [47] -- \citealt{Grankin2013}); column 16 lists the reason, where applicable, for a star's exclusion from the final analysis (a -- not fully convective (see equation \ref{tcore} and Section \ref{AMtheory}), b -- star lies outside the limits imposed in isochronal fitting (see Section \ref{massage}), c -- no optical photometry available to calculate $L_{\star}$ (see Section \ref{Lbol_section}), d -- $P>15\,$days (see Section \ref{Prot_section}), e -- $j_{\star}>j_{\rmn{crit}}$ (see Section \ref{Prot_section}), f -- no reliable rotation period available, g -- able to calculate $j_{\star}$ but not able to classify source as II or III).}
  \begin{tabular}{@{}lccccccccccccccccc@{}}
  \hline
SIMBAD & $\rmn{SpT}$ & $\sigma\left(\rmn{SpT}\right)$ & $T_{\rmn{eff}}$ & $V$ & $B$--$V$ & $I_{\rmn{c}}$ & $V$--$I_{\rmn{c}}$ & $\log\left(L_{\star}\right)$ & $M_{\star}$ & Age & $R_{\star}$ & Period &  Class & Refs & Notes\\ 
  & & & $\left(\rmn{K}\right)$ & & & & & $\left(\rmn{L_{\odot}}\right)$ & $\left(\rmn{M_{\odot}}\right)$ & $\left(\rmn{Myr}\right)$ & $\left(\rmn{R_{\odot}}\right)$ & $\left(\rmn{days}\right)$ & & &\\
(1) & (2) & (3) & (4) & (5) & (6) & (7) & (8) & (9) & (10) & (11) & (12) & (13) & (14) & (15) & (16) \\
 \hline
HD 282600	& $\rmn{K}2$ & $1$ &	$4760$	&	$10.72$	&	$1.62$	&	---	&	---	&	$0.943$	&	$2.28_{-0.57}^{+0.34}$	&	$0.90_{-0.40}^{+0.70}$	&	$4.36\pm0.63$	&	---	&	---	&	21 & a,f \\
HD 282624	& $\rmn{G}8$ & $2$ &	$5210$	&	$9.15$	&	$0.89$	&	$8.10$	&	$1.05$	&	$0.823$	&	$2.21_{-0.25}^{+0.21}$	&	$3.10_{-1.00}^{+1.80}$	&	$3.16\pm0.42$	&	$2.661$	&	II	&	30,11,46,41 & a \\
RY Tau		& $\rmn{K}1$ & $1$ &	$4920$	&	$10.22$	&	$1.03$	&	$8.80$	&	$1.42$	&	$0.729$	&	$2.16_{-0.16}^{+0.15}$	&	$2.10_{-0.80}^{+1.30}$	&	$3.19\pm0.42$	&	$5.6$	&	II	&	41,11,3,29,42 & a \\
HD 283572	& $\rmn{G}5$ & $1$ &	$5500$	&	$9.03$	&	$0.81$	&	$8.10$	&	$0.93$	&	$0.824$	&	$1.98_{-0.25}^{+0.19}$	&	$5.00_{-1.20}^{+2.60}$	&	$2.84\pm0.35$	&	$1.55$	&	III	&	42,11,5,41 & a \\
HD 283782	& $\rmn{K}1$ & $2$ &	$4920$	&	$9.62$	&	$0.84$	&	$8.58$	&	$1.04$	&	$0.580$	&	$1.95_{-0.15}^{+0.13}$	&	$3.10_{-1.10}^{+1.40}$	&	$2.68\pm0.34$	&	---	&	---	&	15,21 & a,f \\
HD 30171	& $\rmn{G}5$ & $1$ &	$5500$	&	$9.26$	&	$0.75$	&	$8.36$	&	$0.9$	&	$0.702$	&	$1.74_{-0.17}^{+0.23}$	&	$7.10_{-2.20}^{+2.40}$	&	$2.47\pm0.30$	&	$1.104$	&	III	&	16,21,47,41 & a \\
HD 285281	& $\rmn{K}1$ & $2$ &	$4920$	&	$12.03$	&	$0.94$	&	$9.69$	&	$1.09$	&	$0.395$	&	$1.69_{-0.15}^{+0.14}$	&	$4.90_{-1.60}^{+2.10}$	&	$2.17\pm0.27$	&	$1.1683$ &	---	&	15,21,47 & a,g \\
GM Aur		& $\rmn{K}3$ & $1$ &	$4550$	&	$10.21$	&	$1.19$	&	$9.12$	&	$2.34$	&	$0.872$	&	$1.69_{-0.53}^{+0.53}$	&	$0.60_{-0.20}^{+0.50}$	&	$4.39\pm0.66$	&	$12$	&	II	&	25,11,24,41 & --- \\
HD 286178	& $\rmn{K}1$ & $2$ &	$4920$	&	$10.30$	&	$0.95$	&	$9.13$	&	$1.08$	&	$0.385$	&	$1.68_{-0.15}^{+0.13}$	&	$4.90_{-1.60}^{+2.40}$	&	$2.14\pm0.27$	&	$1.72$	&	---	&	15,21,33,47 & a,g\\
HD 283641	& $\rmn{K}0$ & $2$ &	$5030$	&	$11.34$	&	$1.32$	&	---	&	---	&	$0.408$	&	$1.67_{-0.16}^{+0.15}$	&	$5.70_{-1.50}^{+2.30}$	&	$2.11\pm0.25$	&	---	&	---	&	15,47 & a,f \\
V1110 Tau	& $\rmn{K}0$ & $1$ &	$5030$	&	$10.09$	&	$0.89$	&	---	&	---	&	$0.376$	&	$1.62_{-0.16}^{+0.16}$	&	$6.30_{-1.70}^{+2.70}$	&	$2.03\pm0.25$	&	$3.039$	&	III	&	42,38,47,29 & a \\
V1298 Tau	& $\rmn{K}1$ & $2$ &	$4920$	&	$10.38$	&	$0.88$	&	$9.36$	&	$1.02$	&	$0.256$	&	$1.51_{-0.17}^{+0.13}$	&	$6.90_{-2.20}^{+3.60}$	&	$1.85\pm0.23$	&	$2.86$	&	---	&	15,21,47 & a,g \\
HD 285957	& $\rmn{K}1$ & $2$ &	$4920$	&	$10.72$	&	$0.93$	&	$9.60$	&	$1.12$	&	$0.222$	&	$1.46_{-0.15}^{+0.14}$	&	$7.70_{-2.60}^{+3.30}$	&	$1.78\pm0.22$	&	$3.0789$ &	---	&	15,21,47 & a,g \\
HD 282630	& $\rmn{K}0$ & $2$ &	$5030$	&	$10.85$	&	$1.02$	&	$9.68$	&	$1.17$	&	$0.232$	&	$1.43_{-0.17}^{+0.14}$	&	$8.90_{-2.20}^{+4.60}$	&	$1.72\pm0.21$	&	$2.2393$ &	III	&	15,11,46,41 & a \\
HD 281691	& $\rmn{K}1$ & $2$ &	$4920$	&	$10.65$	&	$0.85$	&	$9.60$	&	$1.05$	&	$0.178$	&	$1.41_{-0.16}^{+0.13}$	&	$8.60_{-2.70}^{+3.90}$	&	$1.69\pm0.21$	&	$2.662$	&	---	&	15,21,47 & a,g \\
HD 31281	& $\rmn{G}1$ & $2$ &	$5970$	&	$9.22$	&	$0.62$	&	---	&	---	&	$0.571$	&	$1.37_{-0.07}^{+0.10}$	&	$15.00_{-3.00}^{+3.50}$	&	$1.80\pm0.21$	&	---	&	---	&	15,47 & a,f \\
V1299 Tau	& $\rmn{G}3$ & $2$ &	$5740$	&	$9.33$	&	$0.61$	&	$8.61$	&	$0.72$	&	$0.506$	&	$1.36_{-0.09}^{+0.14}$	&	$14.00_{-3.50}^{+4.00}$	&	$1.81\pm0.22$	&	$0.816$	&	---	&	15,21,47 & a,g \\
V1072 Tau	& $\rmn{K}0$ & $2$ &	$5030$	&	$10.34$	&	$0.79$	&	$9.45$	&	$0.89$	&	$0.174$	&	$1.34_{-0.13}^{+0.16}$	&	$10.50_{-2.80}^{+5.00}$	&	$1.61\pm0.19$	&	$2.74$	&	III	&	15,11,24,29 & a \\
V1319 Tau	& $\rmn{G}8$ & $2$ &	$5210$	&	$10.26$	&	$0.65$	&	$9.38$	&	$0.88$	&	$0.205$	&	$1.30_{-0.16}^{+0.13}$	&	$13.00_{-3.30}^{+6.50}$	&	$1.55\pm0.18$	&	$0.736$	&	---	&	15,21,47 & a,g \\
V1079 Tau	& $\rmn{K}3$ & $1$ &	$4550$	&	$12.41$	&	$1.37$	&	$10.79$	&	$1.62$	&	$-0.018$ &	$1.26_{-0.18}^{+0.09}$	&	$7.00_{-3.50}^{+6.00}$	&	$1.58\pm0.24$	&	$5.85$	&	II	&	40,11,32,41 & a\\
HD 284266	& $\rmn{K}0$ & $2$ &	$5030$	&	$10.56$	&	$0.73$	&	$9.68$	&	$0.88$	&	$0.082$	&	$1.23_{-0.12}^{+0.15}$	&	$13.50_{-3.80}^{+6.50}$	&	$1.45\pm0.17$	&	$1.812$	&	---	&	15,21,47 & a,g\\
CW Tau		& $\rmn{K}3$ & $1$ &	$4550$	&	$13.34$	&	$1.37$	&	$11.42$	&	$1.92$	&	$-0.083$ &	$1.21_{-0.11}^{+0.09}$	&	$9.00_{-4.50}^{+7.00}$	&	$1.46\pm0.22$	&	$8.2$	&	II	&	11,32,41 & a\\
HD 285840	& $\rmn{K}1$ & $2$ &	$4920$	&	$10.81$	&	$0.82$	&	---	&	---	&	$0.022$	&	$1.21_{-0.14}^{+0.14}$	&	$13.50_{-4.30}^{+8.00}$	&	$1.41\pm0.18$	&	$1.561$	&	---	&	15,47 & a,g \\
HD 285372	& $\rmn{K}3$ & $2$ &	$4550$	&	$11.69$	&	$1.07$	&	$10.43$	&	$1.26$	&	$-0.098$ &	$1.20_{-0.11}^{+0.08}$	&	$9.50_{-4.10}^{+6.50}$	&	$1.44\pm0.20$	&	$0.574$	&	---	&	15,21,31 & a,g \\
HD 284496	& $\rmn{K}0$ & $2$ &	$5030$	&	$10.81$	&	$0.83$	&	---	&	---	&	$0.015$	&	$1.14_{-0.11}^{+0.15}$	&	$17.50_{-6.50}^{+7.00}$	&	$1.34\pm0.16$	&	$2.7136$ &	---	&	15,38,47 & a,g\\
\hline
\end{tabular}
\label{TAUtable}
\end{minipage}
\end{table}
\end{landscape}

\section{Results and discussion}\label{results}
\begin{figure*}
\centering
\includegraphics[width=7in]{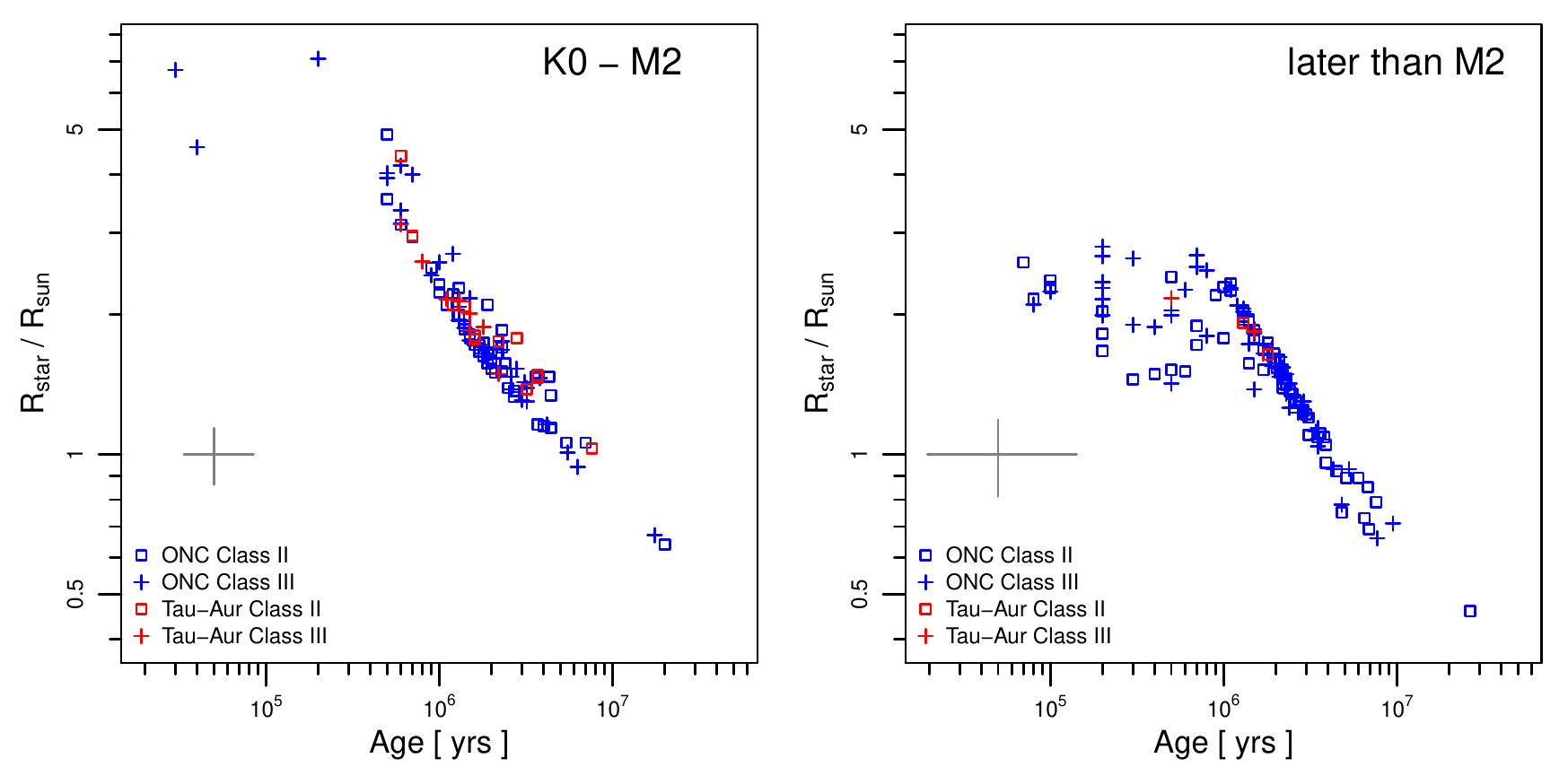}
\centering
\caption{Evolution of the stellar radius of Class II (open squares) and Class III (crosses) stars in the ONC (blue) and Taurus-Auriga (red). The left and right panels show the contraction rates for the high and low mass samples, respectively. An average error bar is located in the bottom left of each plot for reference. We observe a consistent contraction rate in both Class II and Class III objects. Our fitted gradients are presented in Table \ref{rstar_fits} and are slightly steeper than, but in rough agreement with, those expected from purely theoretical considerations of contraction on a Hayashi track.}
\label{contraction}
\end{figure*}

Tables \ref{ONCtable} and \ref{TAUtable} display the data gathered for all members of the ONC and Taurus-Auriga, respectively, that have not previously been identified as binary or multiple (Section \ref{multiplicity}) and for which a spectral type was available in the literature (see Section \ref{membership} for details on ONC membership). Rotation periods found to be affected by beats and harmonics (Section \ref{Prot_section}) are not included. 

As outlined in Section \ref{Prot_section}, we applied several cuts to these data. All stars with (i) rotation periods longer than $15\,$days, (ii) isochronal ages greater than their individual $t_{\rmn{core}}$, or (iii) with $j_{\star}/j_{\rmn{crit}} > 1$ were removed from the analysis (see Sections \ref{massage} and \ref{Prot_section} for details), although they remain included in Tables \ref{ONCtable} and \ref{TAUtable}.

Using equation (\ref{Jstar}), we calculated $j_{\star}$ for all stars for which we had a measured rotation period, stellar radius and an estimate of its age. In total, we were able to calculate $j_{\star}$ for $352$ and $32$ fully convective stars within the ONC and Taurus-Auriga, respectively. Of these, $226$ ONC and $24$ Taurus-Auriga stars were able to be classified as Class II or Class III. We imposed a cut to the data at a spectral type of M2 in order to study our low mass and high mass fully convective PMS stars separately. The final classified samples consisted of $91$ ONC and $20$ Taurus-Auriga stars of spectral types K0 to M2 together with a further $135$ ONC and $4$ Taurus-Auriga stars of spectral type later than M2. These formed our high mass and low mass samples, respectively. 

Before considering how $j_{\star}$ evolves with age for the various samples in Section \ref{am_evol_section}, we first consider its expected time evolution based on theoretical considerations in Section \ref{time_evol}.

\subsection{Evolution of specific angular momentum during PMS contraction: theory}\label{time_evol}
It is clear from equation (\ref{Jstar}) that, as $j_{\star}\propto R_{\star}^{2}/P$, the specific AM evolution of a PMS star depends on the stellar contraction and how the stellar rotation period evolves with time. We consider these quantities in turn. For a contracting fully convective polytropic PMS star, descending a Hayashi track in the HR diagram ($T_{\rmn{eff}}\approx \rmn{const.}$), it is straightforward to show that 
\begin{equation}\label{contraction_eq}
	R_{\star}\propto t^{-1/3}
\end{equation}
 (e.g. \citealt{Lamm2005}).

Fig. \ref{contraction} shows the rate of stellar contraction in our ONC and Taurus-Auriga samples. We used the numerical recipe FITEXY routine in IDL to produce a minimum-$\chi^{2}$ fit to the linear relation
\begin{equation}\label{chisq_rstar}
 \log (R_{\star}) = -\beta_{1} \log (t) + \gamma_{1}.
\end{equation}
This routine can account for symmetric heteroscedastic errors in both $R_{\star}$ and age. However, due to the method of their estimation, the errors in stellar age are not symmetric (Section \ref{massage}). For each value of stellar age, we adopt the maximum of its lower and upper bounded error for the minimum--$\chi^{2}$ fitting procedure. The results of this analysis are presented in Table \ref{rstar_fits}. Under the assumption that the radius and age spreads that we observe in our ONC and Taurus-Auriga samples are real (see Section \ref{radiispread}), we find that the rate of stellar contraction observed in our ONC and Taurus-Auriga samples is steeper than, but in rough agreement with, that expected from equation (\ref{contraction_eq}). Taking $\beta_{1}=1/3$ (equation \ref{contraction_eq}), we would expect $j_{\star}$ to evolve as $j_{\star}\propto t^{-2/3}P^{-1}$.

If, over a timescale of a few Myr, the rotation period of a star varies, on average, as a simple power law of the form
\begin{equation}\label{spin_evol}
	P\propto t^{n},
\end{equation}
 then 
\begin{equation}
	j_{\star}\propto t^{{-2/3} -n}.
\end{equation}
There then exist three scenarios: $n=0$ corresponds to a star that is evolving at a constant rotation rate; $n>0$ to a star that is spinning up; and $n<0$ to a star that is spinning down. The evolution of the rotation period and, therefore, of $j_{\star}$ will differ for Class II and Class III stars with the former being driven by the astrophysics of the star-disc interaction. Assuming that, during the Class II phase, a star is locked to its disc -- accreting and contracting without spinning up, with the surface rotation rate fixed to the Keplerian rotation rate at the disc truncation radius, a common assumption of PMS rotational evolution models (e.g. \citealt{Gallet2013}) -- then $n=0$. Thus, we would expect $j_{\star}$ to reduce with age as $j_{\star}\propto t^{-2/3}$.

Class III stars, which have lost their accretion discs, would conserve AM as they contract such that $j_{\star}=\rmn{const.}$ (neglecting the likely small loss of AM in the stellar wind). Therefore, for Class III stars, we expect $n=-2/3$ such that they spin up as $P\propto t^{-2/3}$ as they continue their gravitational contraction. However, as we discuss in the following subsection, this is not what we observe. Instead, assuming the inferred luminosity spreads for the ONC and Taurus-Auriga are indicative of real age spreads (see Section \ref{agespread}), our results suggest that $j_{\star}\propto t^{-\beta_{2}}$ with $\beta_{2}\approx 2$--$2.5$ for both Class II and Class III sources.

\subsection{Evolution of specific angular momentum during PMS contraction: observations}\label{am_evol_section}
\begin{figure*}
 \centering
 \includegraphics[width=7in]{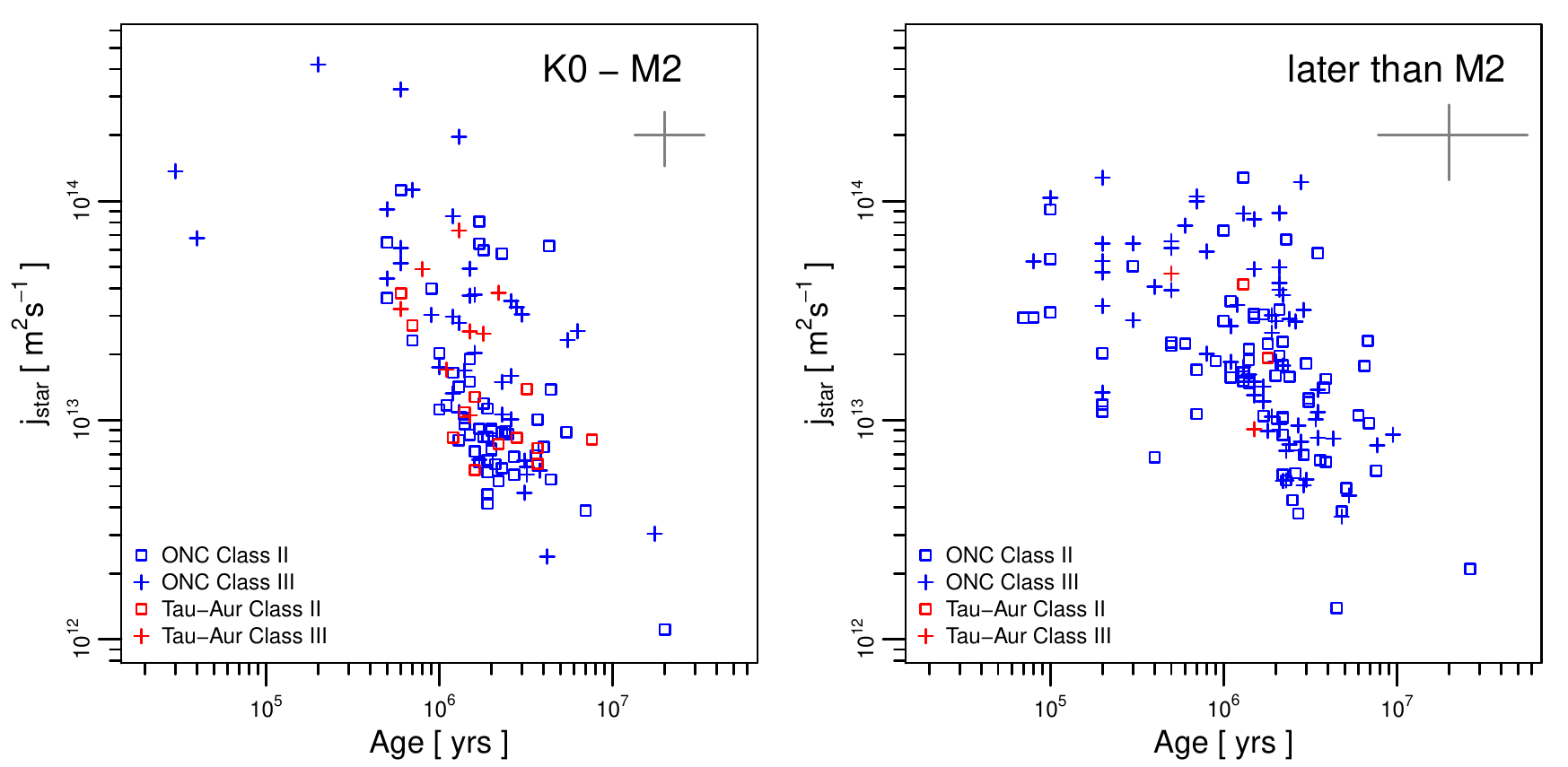}
 \centering
 \caption{Evolution of $j_{\star}$ for the high mass (left panel) and low mass (right panel) Class II and Class III samples in the ONC and Taurus-Auriga. The coloured symbols have the same meaning as in Fig. \ref{contraction}. An average error bar is included in the top right of both plots for reference. We observe a reduction of $j_{\star}$ with increasing stellar age in both of the Class II and Class III samples. The results of a Spearman rank correlation test indicate that we can reject the null hypothesis of zero correlation at statistically significant levels. The results of these correlation tests are presented in Table \ref{AM_fits}.}
 \label{AMevol_plot}
\end{figure*}

Fig. \ref{AMevol_plot} shows the calculated $j_{\star}$ plotted against stellar age for the high mass and low mass samples. We check for the presence of a correlation using a Spearman rank correlation test and present the results of this in Table \ref{AM_fits} for the different masses and classifications. Due to the comparatively low size of the Taurus-Auriga samples, we consider the ONC sample alone and compare it to the combined ONC and Taurus-Auriga sample. The result of combining the Taurus-Auriga and ONC samples does not alter the outcome of the correlation tests significantly, suggesting a consistency between the results in the two regions. 

We consider the evolution of specific stellar AM, $j_{\star}\propto t^{-\beta_{2}}$, in its logarithmic form and fit the linear relation 
\begin{equation}\label{chisq_am}
 \log (j_{\star}) = - \beta_{2} \log (t) + \gamma_{2}
\end{equation}
 using the numerical recipe FITEXY routine in IDL. As in Section \ref{time_evol} with equation (\ref{chisq_rstar}), we used the maximum of the upper and lower bounded errors on the stellar ages in this fitting procedure. The values of $\beta_{2}$, resulting from the fits to our high mass and low mass Class II and Class III samples, are displayed in Table \ref{AM_fits}. We find that $j_{\star}$ decreases with age for both Class II and Class III PMS stars. Furthermore, we find consistent values of $\beta_{2}$ for the Class II and Class III high mass and low mass samples with $\beta_{2}\approx 2$--$2.5$.

\begin{table}
 \centering
 \caption{Results of minimum-$\chi^{2}$ fitting to equation (\ref{chisq_rstar}) for (i) the ONC sample alone, and (ii) the combined ONC and Taurus-Auriga samples. Column $1$ lists the sample name; columns $2$ and $3$ list the value of $\beta_{1}$ for the high and low mass samples, respectively.}
 \begin{tabular}{@{}lcc@{}}
 \hline
Sample & \multicolumn{2}{c}{minimum-$\chi^{2}$ $\beta_{1}$} \\
 & High mass & Low mass\\
(1) & (2) & (3) \\
\hline
ONC Class II & $0.53\pm0.09$ & $0.42\pm0.07$ \\
ONC Class III & $0.53\pm0.08$ & $0.56\pm0.14$ \\
ONC $\&$ Tau Class II & $0.53\pm0.08$ & $0.42\pm0.07$ \\
ONC $\&$ Tau Class III & $0.53\pm0.07$ & $0.56\pm0.14$ \\
\hline
\end{tabular}
\label{rstar_fits}
\end{table}

\begin{figure*}
 \centering
 \includegraphics[width=7in]{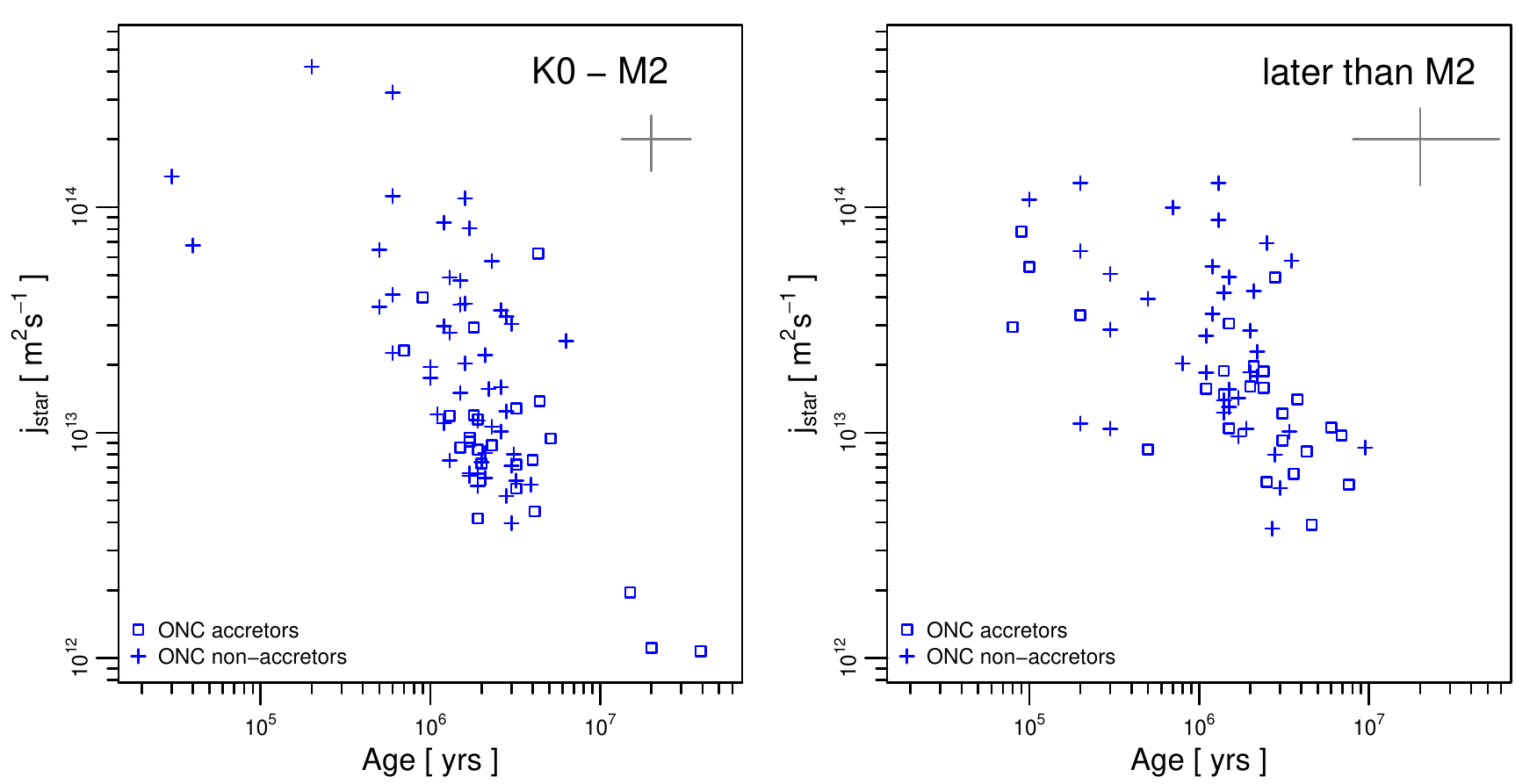}
 \centering
 \caption{As Fig. \ref{AMevol_plot} but for ONC sources identified as accreting and non-accreting rather than Class II and III, respectively. An average error is included in the upper right of each plot for reference. The decrease of $j_{\star}$ is recovered for the low mass Class II and Class III samples and the high mass Class III sample. However, we fail to recover a statistically significant correlation between $j_{\star}$ and age for the high mass Class II sample. This difference is due, in part, to the smaller number of stars identified as accreting compared to disc-hosting. The results of our statistical analysis are presented in Table \ref{AM_fits}.}
 \label{AMevol_acc}
\end{figure*}

We find similar results if we separate our ONC sample by accretion indicators rather than disc indicators (Fig. \ref{AMevol_acc} and Table \ref{AM_fits}). We use the equivalent width of the Ca II $8542\,\rmn{\AA}$ line (one of the Ca II IR triplet lines), $\rmn{EW\left(CaII\right)}$, from \citet{Hillenbrand1998} to distinguish between accretors and non accretors as it has only a weak dependence on spectral type \citep{Hillenbrand1998, White2003}. We identified accretors as having $\rmn{EW\left(CaII\right)} < -1\,\rmn{\AA}$ \citep{Hillenbrand1998} and non-accretors as having $\rmn{EW\left(CaII\right)} > 1\,\rmn{\AA}$, based on the work of \citet{Flaccomio2003}. 

The trend observed in Fig. \ref{AMevol_plot} of decreasing $j_{\star}$ with age is recovered for the low mass accreting and non-accreting samples and the high mass non-accreting sample (Fig. \ref{AMevol_acc}). However, we do not recover a statistically significant correlation between $j_{\star}$ and age for the high mass accreting sample. This is, in part, due to the smaller number of stars in the accreting sample compared to the Class II sample. The results of linear $\chi^{2}$ fitting to equation (\ref{chisq_am}) using the numerical recipe FITEXY routine in IDL are presented in Table \ref{AM_fits} for all accreting and non-accreting samples for which we found a statistically significant correlation. We find values of $\beta_{2}$ consistent with those found when using diagnostics of disc presence rather than accretion. 

Initially, the reduction in $j_{\star}$ with age is surprising, for Class III stars in particular (see Section \ref{time_evol}) which should be conserving AM as they contract. However, as we argue below, it is likely that individual Class III stars are evolving with $j_{\star}\approx \rmn{const.}$ and the observed trend, apparent when considering all the Class III PMS stars in a cluster together, can be naturally explained by Class II PMS stars losing their discs rapidly and at a variety of ages.

\begin{table*}
 \begin{minipage}{176mm}
\centering
 \caption{Results of Spearman rank correlation tests and minimum-$\chi^{2}$ fitting to equation (\ref{chisq_am}) for (i) the ONC sample alone, and (ii) the combined ONC and Taurus-Auriga samples. Column $1$ lists the sample name; columns $2$ and $3$ list the Spearman rank correlation coefficient, $\rho$; columns $4$ and $5$ list the corresponding two-sided probability of finding this value of $\rho$ by chance; columns $6$ and $7$ list the value of $\beta_{2}$ from the minimum-$\chi^{2}$ fit to equation (\ref{chisq_am}) when the correlation is statistically significant (even numbered columns refer to the high mass samples whilst odd numbered columns refer to the low mass samples).}
 \begin{tabular}{@{}lcccccc@{}}
 \hline
Sample & \multicolumn{2}{c}{Spearman $\rho$} & \multicolumn{2}{c}{Spearman $p$--value} & \multicolumn{2}{c}{minimum-$\chi^{2}$ $\beta_{2}$} \\
 & High mass & Low mass & High mass & Low mass & High mass & Low mass\\
(1) & (2) & (3) & (4) & (5) & (6) & (7) \\
\hline
ONC Class II & $-0.54$ & $-0.53$ & $<10^{-4}$ & $<10^{-4}$ & $2.58\pm0.66$ & $1.98\pm0.55$ \\
ONC Class III & $-0.75$ & $-0.67$ & $\ll10^{-4}$ & $\ll10^{-4}$ & $2.15\pm0.41$ & $3.97\pm1.66$ \\
ONC $\&$ Tau Class II & $-0.53$ & $-0.53$ & $<10^{-4}$ & $<10^{-4}$ & $2.34\pm0.53$ & $2.00\pm0.55$ \\
ONC $\&$ Tau Class III & $-0.71$ & $-0.67$ & $\ll10^{-4}$ & $\ll10^{-4}$ & $2.09\pm0.40$ & $4.24\pm1.87$ \\
ONC accretors & $-0.50$ & $-0.68$ & $1.3\times10^{-2}$ & $1.8\times10^{-4}$ & --- & $1.73\pm1.09$ \\
ONC non-accretors & $-0.59$ & $-0.47$ & $<10^{-4}$ & $3.6\times10^{-3}$ & $2.43\pm0.46$ & $4.91\pm3.33$ \\
\hline
\end{tabular}
\label{AM_fits}
\end{minipage}
\end{table*}

\begin{figure*}
 \centering
 \includegraphics[width=7in]{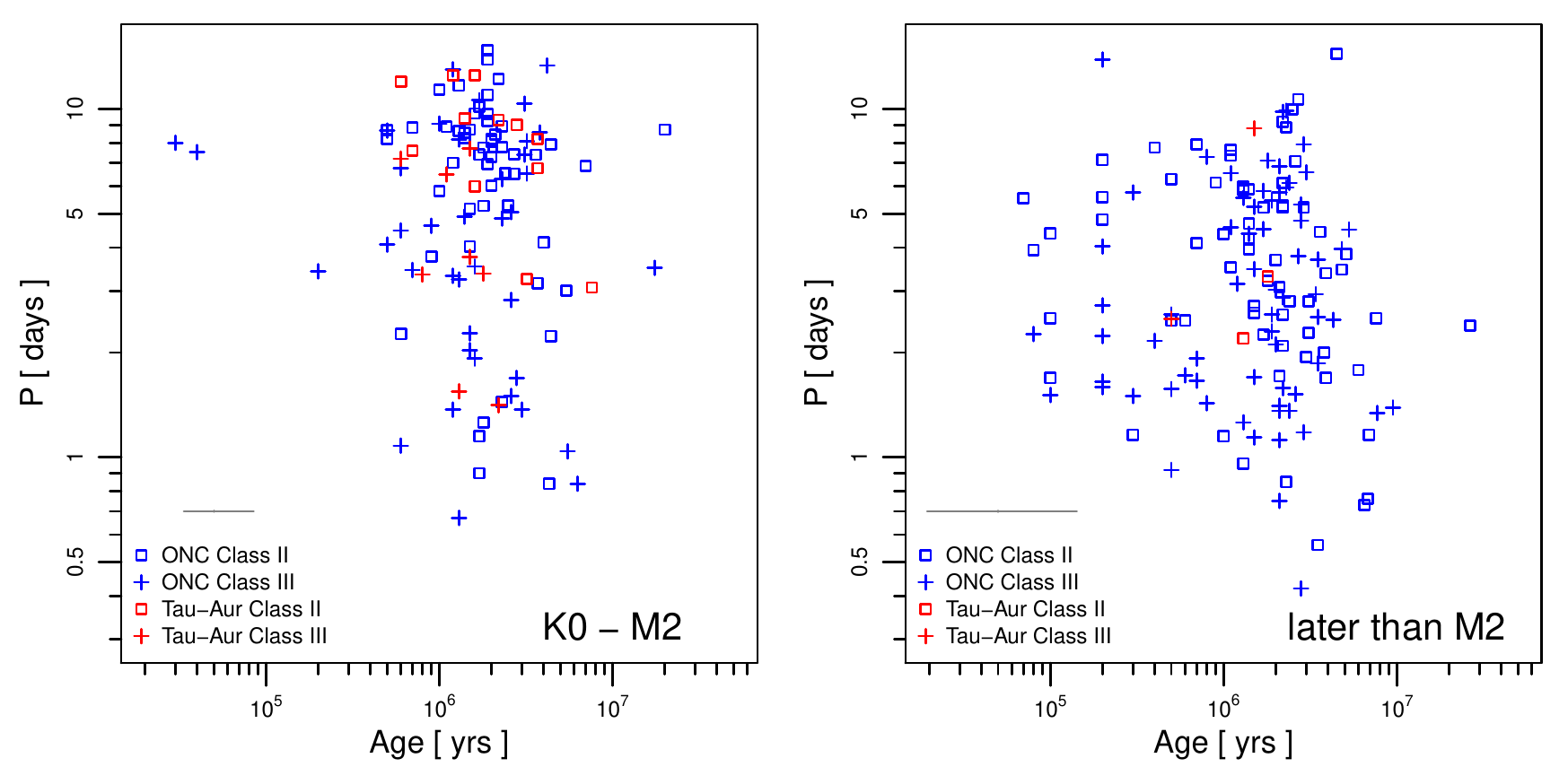}
 \centering
 \caption{Evolution of stellar rotation period for Class II and Class III PMS stars in the ONC and Taurus-Auriga. The left and right panels show the rotational evolution of the high mass and low mass samples, respectively. The coloured symbols have the same meaning as in Fig. \ref{contraction}. We find no statistically significant evidence for a correlation between rotation period and stellar age. The results of the Spearman rank correlation tests we performed are displayed in Table \ref{pevol_fits}.}
 \label{pevol_plot}
\end{figure*}

\subsection{Class III PMS stars}
\begin{table*}
 \begin{minipage}{176mm}
 \centering
 \caption{Results of Spearman rank correlation tests performed on the data in Fig. \ref{pevol_plot} for (i) the ONC sample alone, and (ii) the combined ONC and Taurus-Auriga samples. Column $1$ lists the sample name; columns $2$ and $3$ list the Spearman rank correlation coefficient, $\rho$; columns $4$ and $5$ list the corresponding two-sided probability of finding this value of $\rho$ by chance (even numbered columns refer to the high mass samples while odd numbered columns refer to the low mass samples).}
 \begin{tabular}{@{}lcccc@{}}
 \hline
Sample & \multicolumn{2}{c}{Spearman $\rho$} & \multicolumn{2}{c}{Spearman $p$-value} \\
 & High mass & Low mass & High mass & Low mass \\
(1) & (2) & (3) & (4) & (5) \\
\hline
ONC Class II & $-0.23$ & $-0.27$ & $0.11$ & $0.03$ \\
ONC Class III & $-0.07$ & $0.06$ & $0.67$ & $0.65$ \\
ONC $\&$ Tau Class II & $-0.30$ & $-0.26$ & $0.02$ & $0.03$ \\
ONC $\&$ Tau Class III & $-0.10$ & $0.05$ & $0.51$ & $0.66$ \\
\hline
\end{tabular}
\label{pevol_fits}
\end{minipage}
\end{table*}

\begin{figure*}
 \centering
 \includegraphics[width=5.5in]{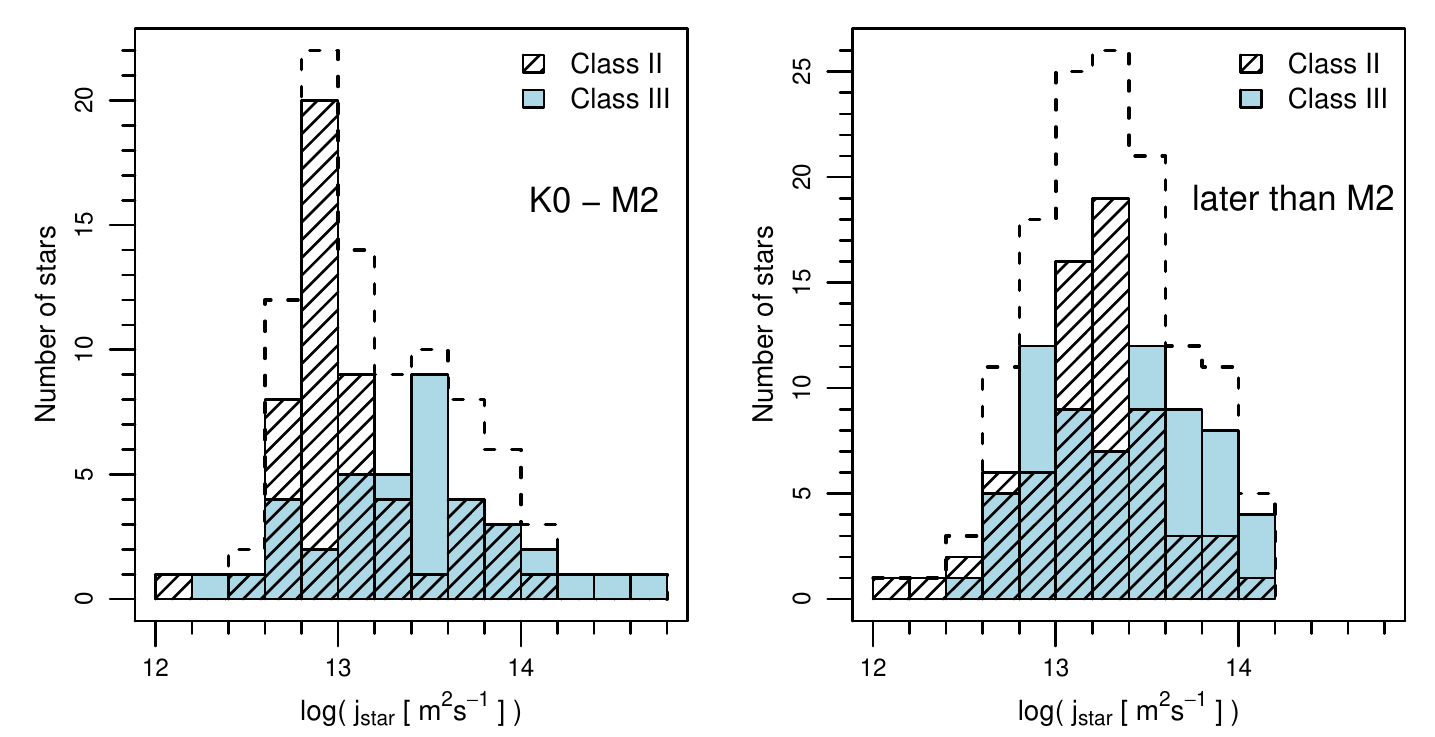}
 \centering
 \caption{Distribution of $j_{\star}$ for high mass (left panel) and low mass (right panel) ONC samples. The full samples are shown as open dashed columns, Class II objects are hatched columns, and Class III objects are shown as blue columns. For both high mass and low mass samples, the Class II PMS stars harbour less $j_{\star}$, on average, than the Class III PMS stars. Double-sided KS tests indicate that the probabilities of the Class II and Class III samples being drawn from the same parent population are $0.00045$ (high mass sample) and $0.016$ (low mass sample).}
 \label{histAM}
\end{figure*}

Once the disc has dispersed, a PMS star undergoing gravitational contraction is expected to conserve AM such that $j_{\star}=\rmn{const.}$ and spin up as $P\propto R_{\star}^{2} \propto t^{-2/3}$ (i.e. $n=-2/3$, see Section \ref{time_evol}). Fig. \ref{pevol_plot} shows the evolution of rotation period for our ONC and Taurus-Auriga samples. We used a Spearman rank correlation test to check for the presence of any correlation and present our results in Table \ref{pevol_fits}. We find our results for Class III high mass and low mass PMS stars are consistent with the null hypothesis where no correlation exists between $P$ and age. However, this does not mean that $n=0$ for these stars and that the specific AM of Class III PMS stars is reducing with age. 

As is visible from the overlap in the ages of Class II and Class III PMS stars in Figs. \ref{AMevol_plot}, \ref{contraction}, and \ref{pevol_plot}, and of accreting and non-accreting sources in Fig. \ref{AMevol_acc}, there is a mixture of stars with and without discs at any given age. This indicates that PMS stars do not lose their discs at the same age. Indeed, analysis of the disc fraction in PMS clusters of various ages has revealed a range of inner disc lifetimes between $1$--$10\,$Myrs (c.f. \citealt{Hillenbrand2005}). Our ONC and Taurus-Auriga samples are consistent with this. Furthermore, double sided Kolmogorov-Smirnov (KS) tests reveal that the probabilities of the Class II and Class III samples being drawn from the same parent population are $0.17$ for the high mass ONC stars, $0.72$ for the low mass ONC stars, and $0.38$ for the high mass Taurus-Auriga stars. This consistency between the ages of Class II and Class III PMS stars highlights how rapid disc dispersal is. 

The location of a Class III PMS star in Fig. \ref{pevol_plot} is a combination of its accretion disc regulated spin evolution, followed by spin up at a rate of $n\approx-2/3$. Without being able to determine the age at which a Class III PMS star lost its disc, there is no way to separate its Class II rotational evolution from its Class III rotational evolution. The predicted spin up of Class III PMS stars during their contraction at constant AM is hidden by the range of disc lifetimes we observe.

This also explains why we see a relationship between $j_{\star}$ and age for the Class III stars consistent with that found for the Class II PMS stars. If all the Class II PMS stars were released from their discs at the same age, we would expect to have $\beta_{2}=0$ (i.e. $j_{\star}=\rmn{const.}$, see Section \ref{time_evol}) for individual stars during the Class III phase, and thus no relation between $j_{\star}$ and age when considering the Class III PMS stars within the cluster as a whole. However, due to the range of disc lifetimes, the location of a Class III star in Fig. \ref{AMevol_plot} (and a non-accreting star in Fig. \ref{AMevol_acc}) is dependent on the efficiency of the AM removal mechanism operating during its disc lifetime combined with the evolution at constant AM following the dispersal of the disc. 

We argue that the difference between the amount of $j_{\star}$ contained within younger and older Class III objects is an artifact of the increasing upper limit of possible disc lifetimes as the star ages. Thus, the younger Class III PMS stars must have had very short disc lifetimes to be observed as such, giving them less time to lose AM during the star-disc interaction (Class II) phase. On the other hand, the older Class III PMS stars do not need to have had such short disc lifetimes. Therefore the younger Class III PMS stars contain more $j_{\star}$ than their older counterparts which, on average, will have had longer disc lifetimes and will, therefore, have spent more time losing $j_{\star}$ before then evolving with constant $j_{\star}$.

\begin{figure}
 \centering
 \includegraphics[width=0.45\textwidth]{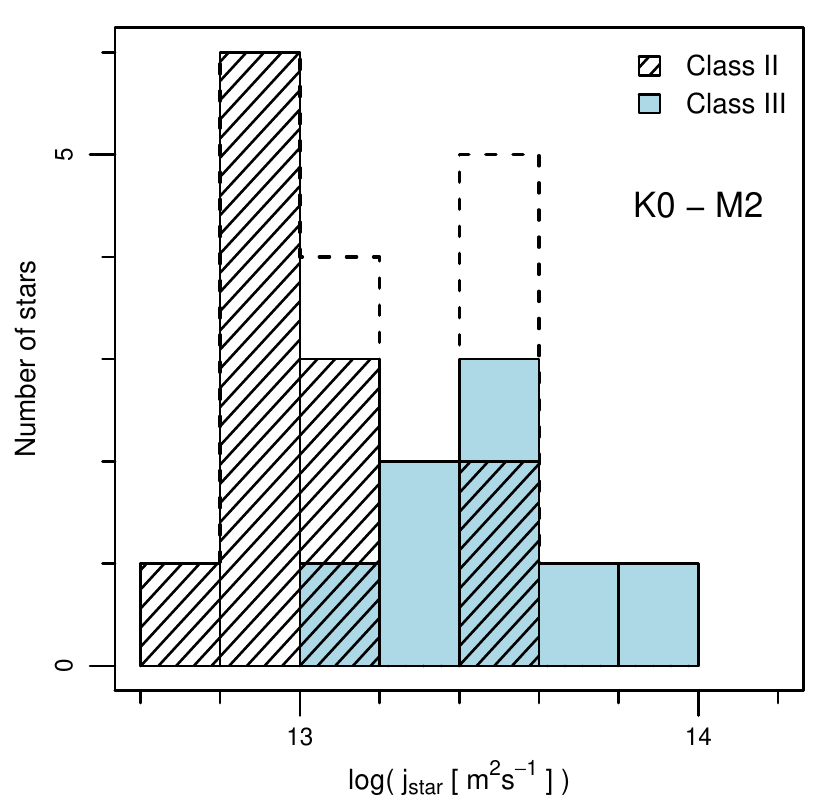}
 \centering
 \caption{As Fig. \ref{histAM} but for the high mass Taurus-Auriga sample. The low mass stars are not shown as there are only $4$ of them. As in Fig. \ref{histAM}, the Class II PMS stars are observed to contain less $j_{\star}$ than the Class III PMS stars. A double-sided KS test reveals that the probability that the Class II and Class III high mass samples are drawn from the same parent population is $0.0086$.}
 \label{histAM_tau}
\end{figure}

This idea is reinforced when we compare the distributions of $j_{\star}$ for Class II and Class III PMS stars within the ONC (Fig. \ref{histAM}) and Taurus-Auriga (Fig. \ref{histAM_tau}). In the high mass and low mass ONC samples and the high mass Taurus-Auriga sample, the Class II objects contain less $j_{\star}$, on average, than the Class III objects. There were not enough data in the low mass Taurus-Auriga sample to perform the same analysis. The mean $j_{\star}$ of the high mass Class II and Class III ONC samples is $1.88\times10^{13}\,\rmn{m^{2}}\,\rmn{s^{-1}}$ and $5.44\times10^{13}\,\rmn{m^{2}}\,\rmn{s^{-1}}$, respectively. Similarly, for the low mass ONC sample, the mean Class II $j_{\star}$ is $2.18\times10^{13}\,\rmn{m^{2}}\,\rmn{s^{-1}}$ whilst the mean Class III $j_{\star}$ is $3.44\times10^{13}\,\rmn{m^{2}}\,\rmn{s^{-1}}$. For the high mass Taurus-Auriga sample, the mean Class II $j_{\star}$ is $1.29\times10^{13}\,\rmn{m^{2}}\,\rmn{s^{-1}}$ and the mean Class III $j_{\star}$ is $3.38\times10^{13}\,\rmn{m^{2}}\,\rmn{s^{-1}}$. A double-sided KS test indicates that the Class II and Class III samples are drawn from the same parent population at probabilities of $0.00045$ (high mass ONC sample), $0.016$ (low mass ONC sample), and $0.0086$ (high mass Taurus-Auriga sample). Thus, the Class II PMS stars which, at any particular age, are still interacting with their discs and losing AM, contain less $j_{\star}$ than Class III PMS stars which have already lost their discs at earlier ages (a Class III PMS star would have been evolving with $j_{\star}=\rmn{const.}$ while $j_{\star}$ was still reducing for the Class II PMS star).  

\subsection{Class II PMS stars}
An accreting (Class II) PMS star may be expected to undergo periods of spin-up and spin-down due to changes in the location of the disc truncation radius, $R_{\rmn{t}}$, relative to the corotation radius, $R_{\rmn{co}}$, over time (e.g. \citealt{Romanova2002, Matt2005}). $R_{\rmn{t}}$ itself is a function of both the magnetic field strength at the inner disc and the mass accretion rate \citep{Konigl1991, Bessolaz2008, Johnstone2014}, both of which are known to vary with time (e.g. \citealt{Donati2011a, Audard2014} and references therein). If we assume, as before, that, on average and over a timescale of a few Myr, the rotation period varies as a power law with $P\propto t^{n}$ then, assuming no prior knowledge of the stellar contraction rate (i.e. $R_{\star}\propto t^{-\beta_{1}}$), it follows from $j_{\star}\propto R_{\star}^{2}/P \propto t^{-\beta_{2}}$ (equations \ref{Jstar} and \ref{chisq_am}) that
\begin{equation}
	\beta_{2}=2\beta_{1}+n.
\end{equation}

In Section \ref{time_evol} we found values of $\beta_{1}$ larger than, but in rough agreement with, purely theoretical considerations of a contracting polytropic star where $\beta_{1}=1/3$. The values of $\beta_{1}$ presented in Table \ref{rstar_fits} are consistent between the different samples. Therefore, we consider $\beta_{1}=1/3$ such that $\beta_{2}$, and therefore the evolution of $j_{\star}$, is dependent only on the value of $n$. In a disc-locked state, a star would spin at the same rate as the Keplerian rotation rate at $R_{\rmn{t}}$ and would evolve with $n=0$ (i.e. at constant $P$). In this case, $\beta_{2}=2/3$ (see Section \ref{time_evol}) such that $j_{\star}\propto t^{-2/3}$. If the net effects of the torques in the star-disc system are such that the star is spinning down (the $n>0$ case), the reduction in $j_{\star}$ with age may be more rapid. Conversely, if the net torques result in the star spinning up (the $n<0$ case), $j_{\star}$ will either decrease (for $-2/3 < n < 0$), remain constant (for $n=-2/3$), or increase (for $n<-2/3$) with age.

The observations discussed in Section \ref{am_evol_section} suggest that, in the ONC and Taurus-Auriga, $j_{\star}$ reduces with age as $j_{\star}\propto t^{-\beta_{2}}$ with $\beta_{2}\approx 2$--$2.5$ for Class II PMS stars. This is a more rapid reduction than is expected if stars are locked to their discs. It suggests that Class II PMS stars may be efficiently spun down during the star-disc interaction phase, despite their contraction and accretion of high AM material from the inner disc. The mechanism by which this can occur is likely some form of outflow (see e.g. \citealt{Zanni2013} and \citealt{Bouvier2013} for up-to-date discussions). 

In apparent contrast to these results, we find no clear correlation between $P$ and age for the Class II stars (see Fig. \ref{pevol_plot}) when considering the entire sample as a whole. However, this does not rule out individual Class II PMS stars being locked to their discs, as there may be a range of disc-locking periods that would depend on variations in the magnetic fields and mass accretion rates across the stars in the sample. If this were the case, we would expect a range of rotation periods among Class II stars, which has long been observed (e.g. \citealt{Herbst2002}; see also Section \ref{rotation_mass}). Additionally, throughout the lifetime of the disc, $n$ may vary such that the torques acting in the star-disc system result in periods of stellar spin-up and stellar spin-down which could also explain the scatter found in the $P$ versus age plots. 

\subsection{Rotation period distributions and the relation between stellar mass and rotation rate}\label{rotation_mass}
\begin{figure*}
 \centering
 \includegraphics[width=5.5in]{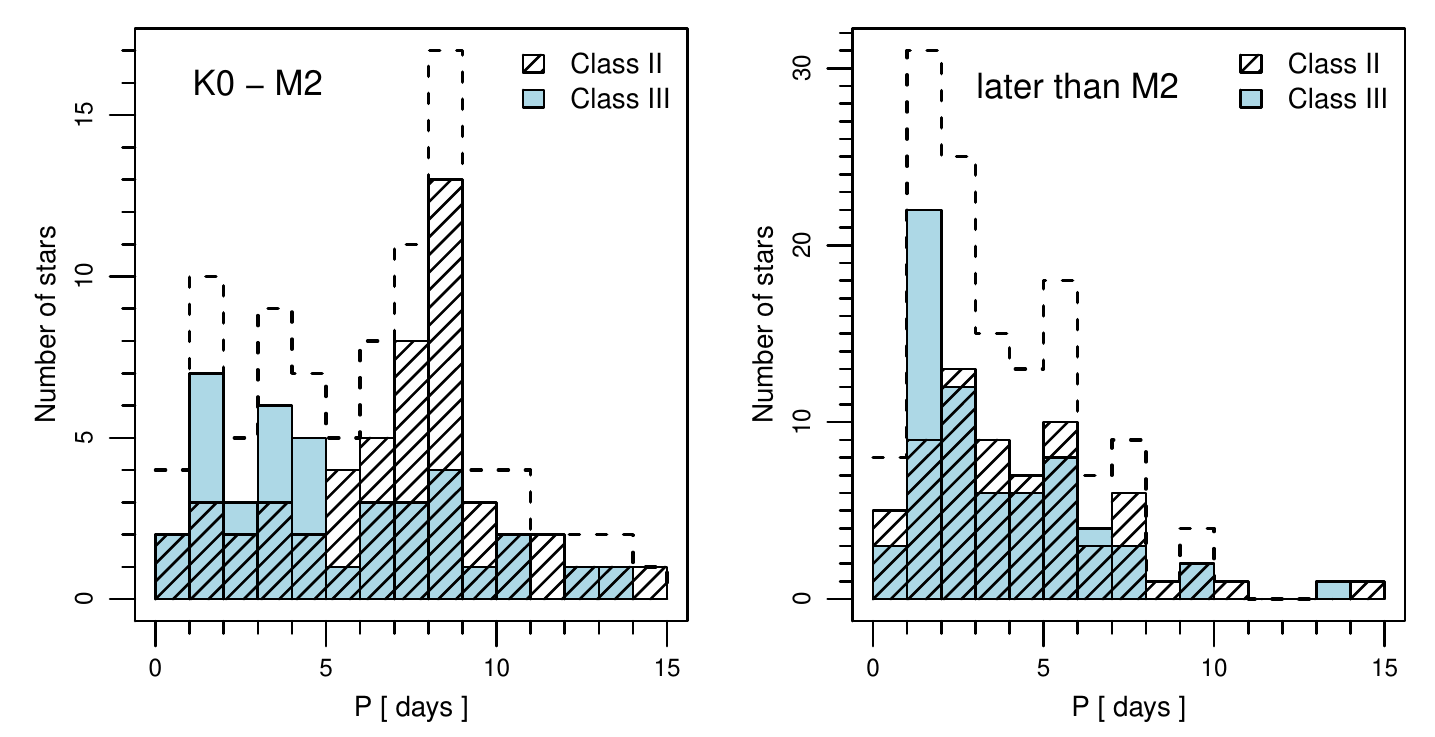}
 \centering
 \caption{Distribution of rotation periods for the high mass (left panel) and low mass (right panel) ONC samples. The full samples are shown as open dashed columns, the Class II objects are hatched columns and the Class III objects are blue columns. The previously observed bimodal distribution is recovered for the high mass sample and the previously observed unimodal distribution is found for the low mass sample. For the high mass stars, the mean rotation periods are $7.07\,$days (Class II) and $5.11\,$days (Class III) and, for the low mass stars, the mean rotation periods are $4.27\,$days (Class II) and $3.57\,$days (Class III). A double-sided KS test indicates that the probability of the Class II and Class III samples being drawn from the same parent population is $0.0027$ (high mass) and $0.16$ (low mass).}
 \label{phist_onc}
\end{figure*}

\begin{figure}
 \centering
 \includegraphics[width=0.45\textwidth]{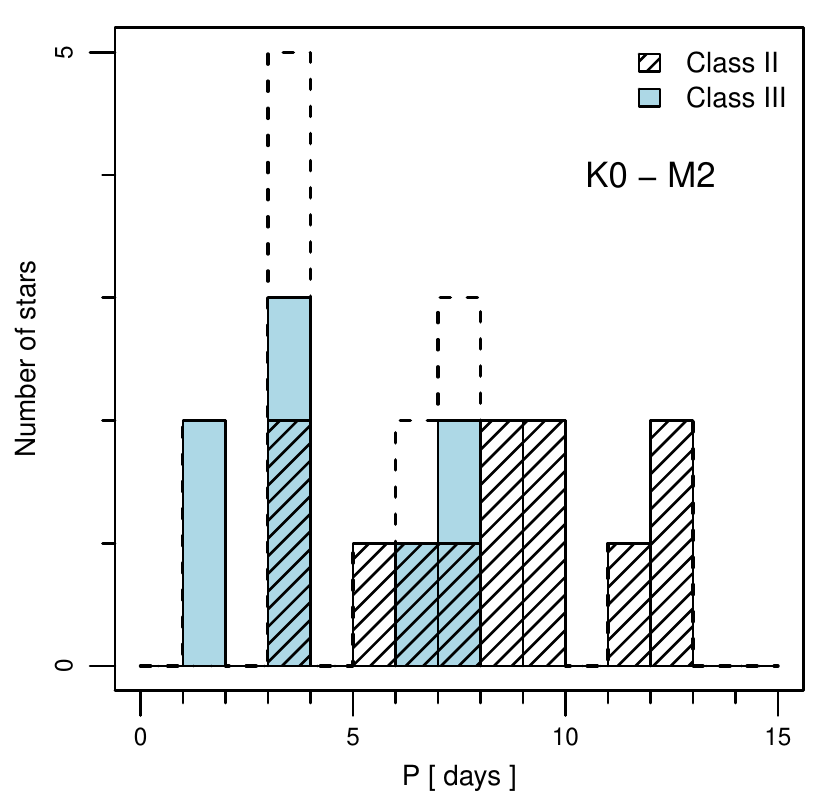}
 \centering
 \caption{Distribution of rotation periods for the high mass Taurus-Auriga sample. The full samples are shown as open dashed columns, the Class II objects are hatched columns and the Class III objects are blue columns. On average, the Class II PMS stars are slower rotators than the Class III PMS stars. The mean rotation periods are $8.30\,$days (Class II) and $4.35\,$days (Class III). However, due to the size of the sample, this result is not statistically significant. A double-sided KS test indicates that the probability of the Class II and Class III high mass samples being drawn from the same parent population is $0.076$.}
 \label{phist_tau}
\end{figure}

In order to observe $n=0$ for the full sample of ONC and Taurus-Auriga stars (Fig. \ref{pevol_plot} and Table \ref{pevol_fits}), we would expect the individual stars to display a range of disc-locking periods. Figs. \ref{phist_onc} and \ref{phist_tau} show the distributions of rotation periods for the ONC and Taurus-Auriga samples, respectively. A range of rotation periods is observed for both the Class II and Class III samples suggesting that, if $n=0$, a range of disc-locking periods do exist. Additionally, we recover the bimodal distribution seen previously for our high mass ONC sample \citep{Herbst2002, CiezaBaliber2007} with the Class II PMS stars rotating at slower rates, on average, than the Class III PMS stars, suggesting an accretion disc regulated AM removal mechanism operates. A double-sided KS test indicates that the high mass Class II and Class III ONC samples are drawn from the same parent population at a probability of $0.0027$. In the comparatively small high mass Taurus-Auriga sample, the bimodality is also visible and the average rotation period of the Class II sample is, again, larger than that of the Class III sample. However, a double-sided KS test does not reveal a statistically significant probability of the Class II and Class III rotation periods are drawn from the same parent population ($0.08$). 

The observed bimodal distribution for the high mass samples is interpreted as indicating a degree of accretion disc regulated rotation during the Class II phase, followed by spin up during the Class III phase. However, we find that not all Class III objects are rapid rotators. It is possible that the slowly rotating Class III sources have only recently been released from their discs and have not yet had chance to spin up. Similarly, the peak of rapid rotators also hosts stars that indicate the presence of a disc. It is possible that these stars have disc truncation radii closer to their photospheres and so are locked to a faster spinning region of the Keplerian disc. The disc truncation radius is related to the mass accretion rate and the dipole component of the large scale stellar magnetic field (e.g. \citealt{Adams2012}) so Class II sources in the peak of rapid rotators could have higher accretion rates and/or weaker dipole components of their magnetic fields. 

We can extend this idea to our low mass ONC sample. We recover the unimodal distribution of rotation periods seen previously for the low mass ONC stars (e.g. \citealt{Herbst2002}). We find that the rotation periods of low mass Class II and Class III objects are consistent. A double-sided KS test reveals a probability of $0.16$ that they are drawn from the same parent population. This could be explained if the lowest mass fully convective stars have disc truncation radii closer to their stellar surfaces than the higher mass fully convective stars as a result of weaker dipole components of their large-scale magnetic fields. \citet{Donati2010a} and \citet{Gregory2012} argue that the lowest mass PMS stars may have complex magnetic fields, which would result in smaller disc truncation radii and, therefore, faster disc-locked stellar spin rates than found for higher mass fully convective stars. Additional observations of the magnetic field topologies of the lowest mass PMS stars are required to confirm this. 

\section{Summary}\label{summary}
We have studied the evolution of $j_{\star}$ in fully convective stars during the Class II and Class III stages of PMS evolution. To do this, we have constructed a consistent sample of PMS stars within the ONC and Taurus-Auriga, gathering rotation periods from the literature and checking for the effects of beats and harmonics. We take into account the recently updated spectral type assignments and new spectral types that have been reported in the literature for the first time (e.g. \citealt{Hillenbrand2013}). Effective temperatures were assigned to these spectral types, and bolometric luminosities were calculated from optical photometry, using intrinsic colours, spectral-type-to-effective-temperature conversions, and bolometric corrections appropriate for $5$--$30\,$Myr old PMS stars \citep{Pecaut2013}. These are an improvement over the typically used main sequence dwarf scales as they take into account the combined effects of the lower surface gravities and spotted surfaces of PMS stars. We used the effective temperatures and bolometric luminosities to calculate stellar radii, under the assumptions that the stars radiate as black-bodies and that the observational uncertainties associated with estimating both of these quantities are not enough to explain the inferred spread in stellar radii evident from the location in the HR diagram. We estimate stellar masses and ages consistently across the entire sample, taking into account individual errors on spectral type, using \citet{Siess2000} PMS evolutionary models. 

With the spectral type updates and our careful removal of rotation period bias, non-members, and known binaries, we recover the bimodal distribution of rotation periods seen previously for the high mass stars in our sample (e.g. \citealt{Attridge1992, Edwards1993, Choi1996, Herbst2000}) as well as the unimodal distribution seen for the low mass stars \citep{Herbst2002, CiezaBaliber2007}. We find that stars with discs are typically slower rotators across all samples. Each sample has a range of rotation periods with the peaks of both rapid and slow rotators populated by both Class II and Class III sources. The slowly rotating Class III PMS stars have probably recently lost their discs while the faster rotating Class IIIs have spun up. If disc-locking operates, the rapidly rotating Class II PMS stars are likely to have larger mass accretion rates and/or weaker magnetic fields than the slower rotating Class IIs. The slower rotation rates of the higher mass fully convective stars compared to the lower mass fully convective stars is most likely due the more complex large-scale magnetic fields of low mass stars, as indicated by the analysis of magnetic field topologies in PMS stars \citep{Donati2010a, Gregory2012}. 

If we assume that the age spreads that we observe in the ONC and Taurus-Auriga are real (see below and Section \ref{radiispread}), we find that $j_{\star}$ reduces with age for both the Class II and Class III PMS stars, with $j_{\star}\propto t^{-\beta_{2}}$ and $\beta_{2}\approx 2$--$2.5$. For Class II stars, this suggests that they are losing angular momentum at a faster rate than would be required for them to be locked to their discs during contraction. Instead, it suggests that they are spinning down due to an efficient angular momentum removal process in the star-disc system. 
Considering the sample as a whole, we do not find any correlation between the rotation period and age. However, we find that Class II stars typically rotate at slower rates, emphasizing that discs do play a role in regulating the rotation of accreting PMS stars. It is likely that a range of disc-locking rotation periods exists due to variations in the mass accretion rate and the magnetic field both in the same star over time (e.g. \citealt{Donati2011a, Audard2014} and references therein) and between the different stars in our sample \citep{Donati2010a, Gregory2012}.

We would expect individual Class III stars to conserve angular momentum as they contract. Instead, we find that, as a cluster sample, $j_{\star}$ reduces with age for Class III PMS stars at roughly the same rate as the Class II sample. On average, Class III PMS stars have higher $j_{\star}$ than Class II stars. This can be explained by Class II stars losing their discs at a variety of ages (indeed, there are a mixture of stars with and without discs at any particular age within our ONC and Taurus-Auriga samples). Then, if we consider two Class II PMS stars with the same initial $j_{\star}$, losing angular momentum at the same rate, the one that loses its disc (and is observed as a Class III PMS star) will evolve with constant $j_{\star}$, whilst the one that retains its disc (and is observed as a Class II PMS star) will continue to lose angular momentum. 

The correlations observed here ultimately depend on the accuracy with which stars can be positioned within the HR diagram. Throughout this study, we have assumed that the spread in stellar luminosities corresponds to a true spread in stellar radii and that this, in turn, corresponds to a true age spread within the ONC and Taurus-Auriga (see Section \ref{radiispread} for a detailed discussion). However, if our assumptions are incorrect and either the stars are coeval or their ages are indicative of differing accretion histories (e.g. \citealt{Littlefair2011}), our conclusions will require further confirmation. Consequently, the results presented here require further examination using different stellar age indicators, preferably independent of stellar radius measurements, or more reliable bolometric luminosity calculations -- something that will improve dramatically when data from the $\itl{Gaia}$ satellite is available. 

\section*{Acknowledgments}
This research has made use of the SIMBAD database, operated at CDS, Strasbourg, France. We wish to thank Stuart Littlefair and Aleks Scholz for helpful discussions, and Nathan Mayne for a detailed reading of the manuscript. CLD's PhD is supported by a Science and Technology Facilities Council (STFC) studentship from the government of the United Kingdom. SGG acknowledges support from the STFC via an Ernest Rutherford Fellowship [ST/J003255/1]. This research has made use of NASA's Astrophysics Data System. This research has made use of the VizieR catalogue access tool, CDS, Strasbourg, France. We thank the referee, William Herbst, for his helpful comments which greatly improved the quality of this paper.

\footnotesize
\bibliographystyle{mn2e}
\setlength{\bibhang}{2.0em}
\setlength{\labelwidth}{0.0em}
\bibliography{angmom}
\normalsize

\label{lastpage}

\end{document}